\documentclass[12pt,english,preprint,authoryear]{elsarticle}
\usepackage[T1]{fontenc}
\usepackage[latin9]{inputenc}
\usepackage{amsmath}
\usepackage{graphicx}
\usepackage{amssymb}

\makeatletter

\usepackage{epsfig}

\journal{Advances in Space Research} 

\usepackage{babel}

\begin{document}
\begin{frontmatter}



\title{WMAP extragalactic point sources as potential Space VLBI calibrators}

%\maketitle

%\author{Katinka Ger\'eb\corauthref{cor}}

\author{Katinka Ger\'eb\corref{cor}}

\address{Department of Astronomy, E\"otv\"os University, P\'azm\'any P. s\'et\'any 1/A,
H-1117 Budapest, Hungary\\
Kapteyn Astronomical Institute, University of Groningen, P.O. Box 800, 9700 AV Groningen, The Netherlands}

\cortext[cor]{Corresponding author}

\ead{gereb@astro.rug.nl}

\author{S\'andor Frey}

\address{F\"OMI Satellite Geodetic Observatory, P.O. Box 585, H-1592 Budapest,
Hungary\\
MTA Research Group for Physical Geodesy and Geodynamics, P.O.
Box 91, H-1521 Budapest, Hungary}

\ead{frey@sgo.fomi.hu}


\begin{abstract}
The point source list of the Wilkinson Microwave Anisotropy Probe
(WMAP) is a uniform, all-sky catalogue of bright sources with flux
density measurements at high (up to 94~GHz) radio frequencies. We
investigated the five-year WMAP list to compile a new catalogue of
bright and compact extragalactic radio sources to be potentially studied
with Very Long Baseline Interferometry at millimeter wavelengths (mm-VLBI)
and Space VLBI (SVLBI). After comparing the WMAP data with the existing
mm-VLBI catalogues, we sorted out the yet unexplored sources. Using
the 41, 61 and 94~GHz WMAP flux densities, we calculated the spectral
indices. By collecting optical identifications, lower-frequency radio
flux densities and VLBI images from the literature, we created a list
of objects which have not been investigated with VLBI at 86 GHz before.
With total flux density at least 1 Jy and declination above $-40^{\circ}$,
we found 37 suitable new targets. It is a nearly 25\% addition to
the known mm-VLBI sources. Such objects are also potentially useful
as phase-reference calibrators for the future Japanese SVLBI mission
ASTRO-G at its highest observing frequency (43~GHz). The phase-referencing
capability of ASTRO-G would allow long integrations and hence better
sensitivity for observing faint target sources close to suitable phase
calibrators in the sky. 
\end{abstract}

\begin{keyword}
Compact extragalactic radio sources \sep Active galactic nuclei \sep
Radio spectrum \sep mm-VLBI \sep Space VLBI


\end{keyword}
\end{frontmatter}

% main text

\section{Introduction}

\label{intro}

The Wilkinson Microwave Anisotropy Probe (WMAP) is a NASA Explorer
mission, launched in June 2001 to make fundamental measurements of
cosmology \citep{Bennett}. Its main aim is to measure the Cosmic
Microwave Background (CMB) temperature anisotropies at five different
frequencies bet\-ween 23 and 94~GHz. The foreground emission -- galactic emission, galactic and extragalactic point sources -- is
a cause of a major complication because it contaminates the CMB maps.
In order to separate the CMB data and the foreground emission, extragalactic
point source catalogues were created for each WMAP data release. Here
we use another, more extended point source catalogue constructed by
\citet{Chen} based on the five-year WMAP sky maps at three different
frequencies (41, 61 and 94~GHz), using a method that favours the
selection of flat-spectrum sources. This catalogue offers an opportunity
to study the population of the brightest flat-spectrum radio sources
in the millimeter wavelength range. Unlike any ground-based instrument
limited by the observable sky at a given geographic location, WMAP
in fact provides a homogeneous all-sky mm wavelength survey.

The technique of Very Long Baseline Interferometry (VLBI) employs
a network of radio telescopes that simultaneously observe the same
radio source. By combining the measurements recorded at the stations
(or transferred directly to a central processing facility in real
time), the angular resolution of the synthesized large radio telescope
is determined by the maximum separation (baseline) between the individual
telescopes. VLBI allows the imaging of bright and compact radio sources,
including radio-emitting active galactic nuclei (AGN) at extremely
high angular resolution. The first successful 89~GHz VLBI observation
on a single baseline was done by \citet{Read}. Since then, several
other VLBI experiments have been made at mm-wavelengths in order to
investigate the physical properties of quasars and other AGN. For
example, using the 100 and 86 GHz observations made with the Coordinated
Millimeter VLBI Array (CMVA) in 1990 and 1993, \citet{Ranta} created
a catalogue of 16 radio sources, to probe the physics of the innermost
jets in AGN. They achieved up to 50 micro-arcsecond ($\mu$as) angular
resolution while imaging 12 sources. \citet{Lobanov} presented 86 GHz
VLBI observations of 28 radio sources. This catalogue contains 26
AGN, the remaining two sources are the center of our Galaxy (Sgr A$^{\ast}$)
and the Cygnus X-3 X-ray binary star.

The catalogue of \citet{Lee} was a great leap forward, containing
127 compact radio sources (88 quasars, 25 BL Lac objects, 11 radio
galaxies, 1 star and 2 unidentified sources). To define the target
list for their survey, \citet{Lee} selected sources which \textit{(i)}
have not been investigated before at 86~GHz with VLBI, \textit{(ii)}
have declination $\delta$ $>$ $-40^{\circ}$ to ensure a good radio
station coverage, and \textit{(iii)} have 86 GHz flux density $S_{86}>0.3$~Jy.
They produced images for 109 objects. Twelve others were also detected
but could not be imaged due to insufficient data.

At lower frequencies (24 and 43~GHz), the United States Naval Observatory
(USNO) maintains an extensive catalogue.%
\footnote{http://rorf.usno.navy.mil/RRFID$\_$KQ%
} It contains AGN that are being studied for assessing their astrometric
suitability for use in a high-frequency celestial reference frame
\citep{Charlot}.

Space VLBI (SVLBI) involves an orbiting radio telescope as one of
the interferometer elements, to increase the baseline lengths and
hence the achievable angular resolution at a given observing frequency,
with respect to what is available from the Earth-based networks. The
first dedicated SVLBI satellite, the Japanese HALCA was launched in
1997 \citep{Hira}. ASTRO-G is a planned Japanese second-generation
SVLBI mission \citep{Tsuboi} currently foreseen to be launched in
2016. The satellite will carry out observations at 8.4, 22 and 43~GHz,
on interferometric baselines several times longer than the Earth diameter.
ASTRO-G will provide higher angular resolution (down to 38 $\mu$as)
and better sensitivity than any SVLBI experiment before. The high
sensitivity is mostly achieved by long integrations well in excess
of the atmospheric coherence time at the ground-based antennas of
the network. This is made possible by the phase-referencing technique
which involves rapid switching between the scientifically interesting
weak target and a nearby strong calibrator. However, the chance for
finding a suitable phase-reference calibrator source, which
is sufficiently close to the target severely decreases at higher
frequencies \citep{Asaki}. For enhancing the scientific gain of the
mission, it is crucial to find as many compact radio AGN at high frequencies
as possible.

It is very important for several reasons to increase the number of
radio sources mapped at mm-wavelengths. From an astrophysical
point of view, these highest-resolution imaging data are essential
to test the physical models of the inner jets, in the region of jet launching near the central supermassive black holes.
By combining ground-based 86 GHz VLBI images with 43 GHz SVLBI data
obtained with baselines larger by a factor of $\sim$2, and thus at
similar angular resolution, it is possible to create detailed spectral
index images. For high-frequency SVLBI, there is also a need to increase
the sky density of bright sources that are compact at $\sim$10~$\mu$as
scales. These could serve as phase-reference sources for observing
nearby fainter targets.

In this paper, we describe our method to identify new potential mm-VLBI
and SVLBI target sources from the WMAP point source catalogue, and
give the list of the brightest 37 sources selected (Sect.~\ref{targets}).
We also give perspectives on finding more sources in the near future
(Sect.~\ref{summary}).

\section{New mm-VLBI targets from the WMAP point source catalogue}

\label{targets}

Which AGN are suitable as new candidates for observing with mm-VLBI?
To answer this question, we compared the WMAP point source list \citep{Chen}
with the existing VLBI catalogues at high frequencies. After the identifications,
we created sub-samples from the already known mm-VLBI sources and
from those which do not appear in earlier VLBI lists. We also made
a cross-indentification with the most extensive 86 GHz VLBI
catalogue to date \citep{Lee}.

\begin{figure}
\begin{centering}
\includegraphics[width=11cm]{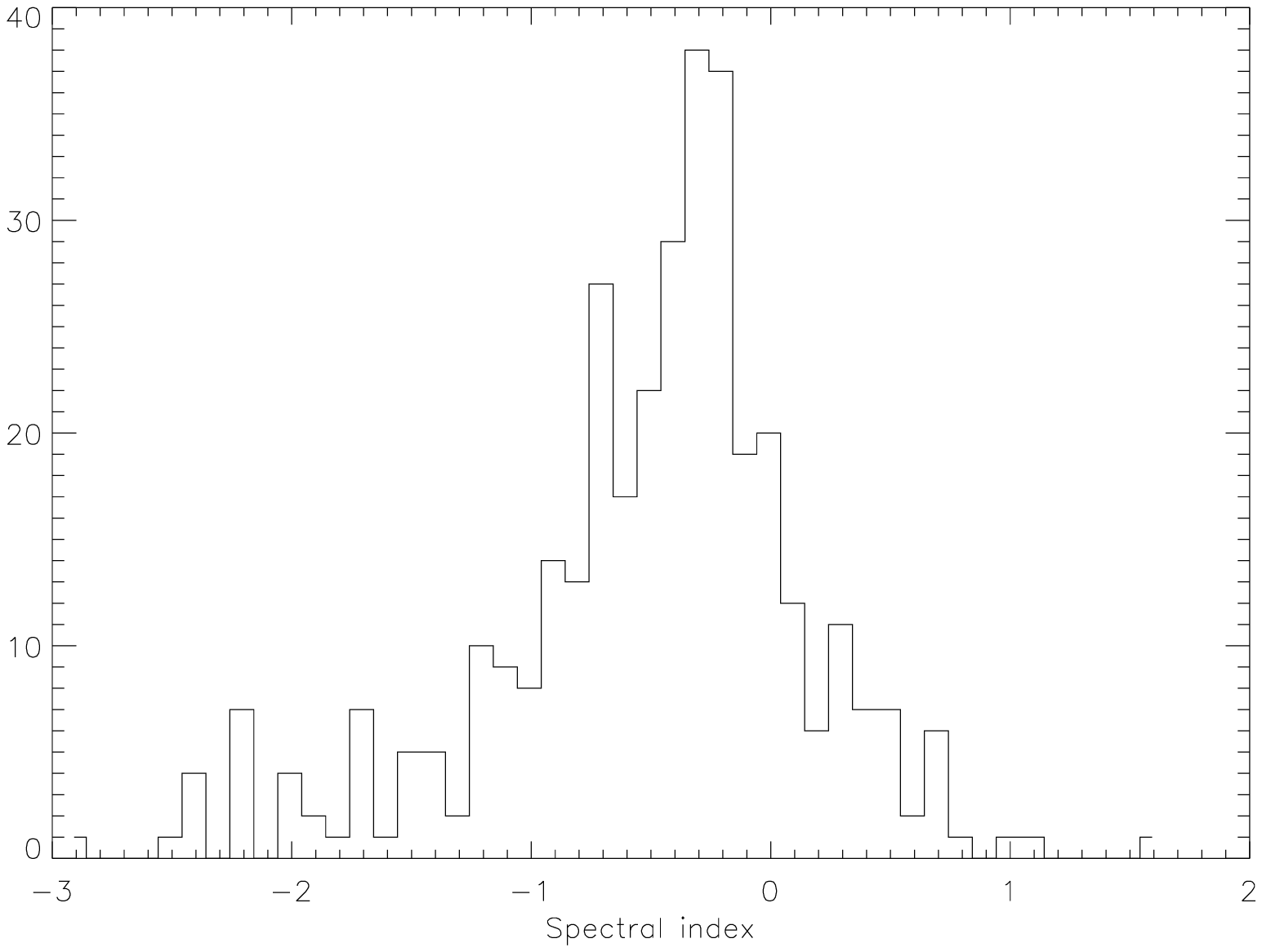}
\includegraphics[width=11cm]{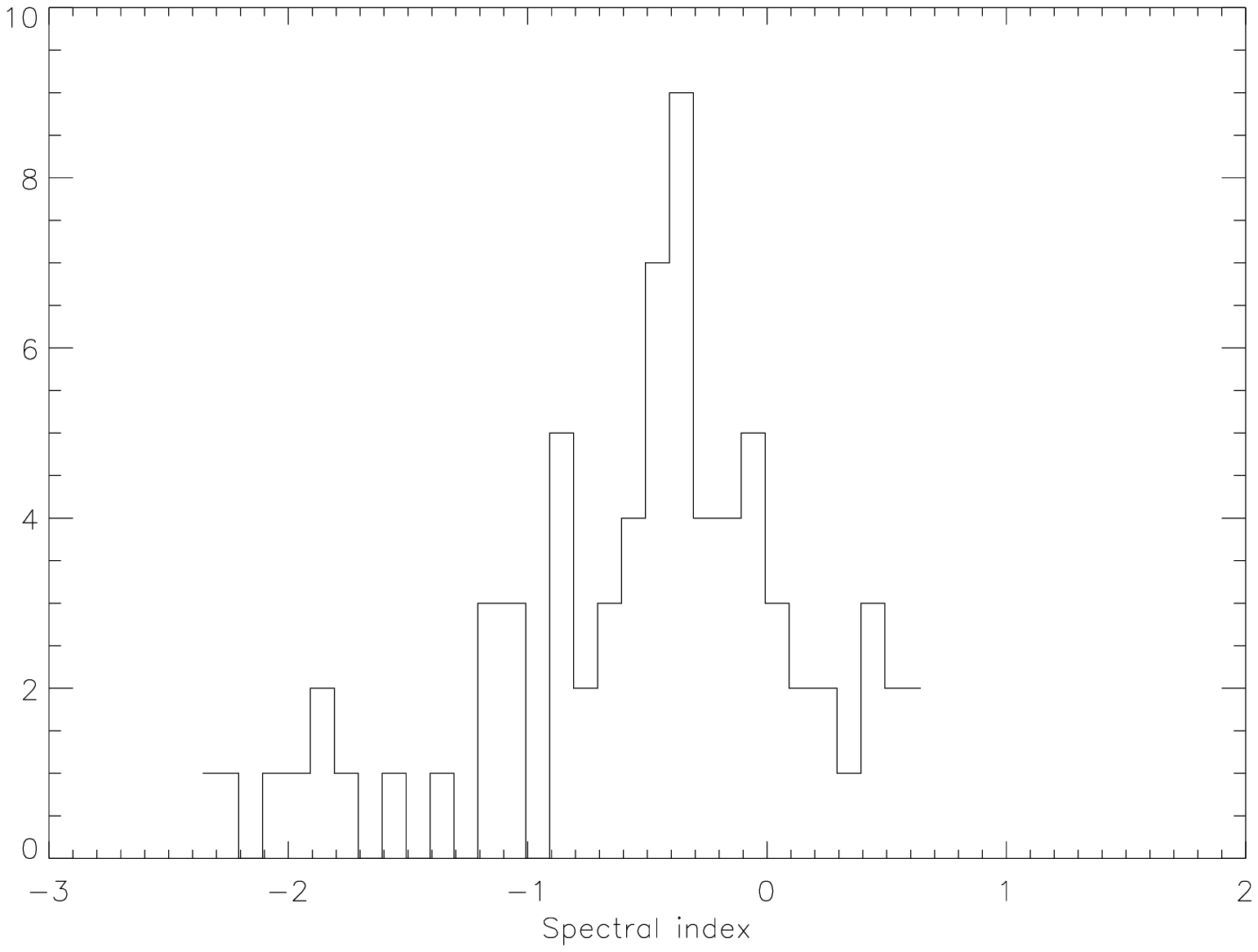}
\par\end{centering}

\caption{Spectral index histograms of the whole WMAP sample \textit{(top)}
and its overlap with the biggest 86-GHz VLBI sample to date \citep{Lee}
\textit{(bottom)}. The two distributions are similar, the WMAP-only
sources have somewhat flatter radio spectra on average. The
median spectral index for the whole WMAP sample is $-0.30$. For the
WMAP sources that are known mm-VLBI objects as well, the median spectral
index is $-0.37$. The median absolute deviations are $0.48$ and $0.37$ respectively.}

\label{histograms} 
\end{figure}

A power-law is usually a good approximation of the continuum radio
spectra in the cm and mm wavelength range. The flux density ($S$)
is the function of the observing frequency ($\nu$) as $S\sim\nu^{\alpha}$,
where $\alpha$ is the spectral index. Using the 41, 61 and 94~GHz
WMAP flux densities, we calculated the radio spectral index for each
source. Note that the point source flux densities are extracted
from the 5-year WMAP temperture maps \citep{Chen} and do not represent
a specific epoch. Any possible flux density variability over the 5-year
interval is avaraged out this way. The spectral indices cannot be
considered as simultaneous. The radio spectra of compact sources
are usually flat (i.e. $\alpha$ $\geq$ $-0.5$) as a result
of the superposition of several small synchrotron self-absorbed components
with different spectral turnover frequencies. For direct comparison
in the VLBI band, we also estimated the source flux densities at 86~GHz,
using the spectral index and the flux density measured in the nearest
WMAP band (94~GHz). Then we compared the spectral index and flux
density distribution of the whole WMAP sample \citep{Chen} and our
sub-samples. It turned out that on average the WMAP catalogue contains
fainter sources with slightly flatter spectra than the earlier VLBI
catalogues. The spectral index histograms of the WMAP and the \citet{Lee}
samples are shown as an example in Fig.~\ref{histograms}.

%%%%%%%%%%%%%%%%%%%% Fig. 1
%

%%%%%%%%%%%%%%%%%%%%

For creating a new list of bright and compact quasars that are suitable
for subsequent mm-VLBI and SVLBI studies, we adopted the selection
method of \citet{Lee}, using the WMAP point source catalogue \citep{Chen}
as the initial sample. First we excluded those WMAP sources that have
already been investigated with 86 GHz VLBI before. Then we applied
1~Jy as the lower limit of total flux density. According
to \citet{Chen}, the sources with $S>1$~Jy are reliable detections
in their WMAP list. However, the number of suitable sources could
be increased if a less conservative lower flux density limit is used. Following \citet{Lee}, only the objects above $-40^{\circ}$
declination were considered since the antennas of the Global Millimeter
VLBI Array (GMVA) are in the northern hemisphere. As a result, we
obtained a list of 38 radio sources. A comparison with the properties
of the \citet{Lee} catalogue sources shows that our new objects are
in general fainter but their spectra are somewhat flatter, and thus
likely sufficiently compact for VLBI detection.

To verify the suitability of our new sources, and to supplement the
new catalogue with other basic data, we searched the NASA/IPAC Extragalactic
Database (NED)%
\footnote{http://nedwww.ipac.caltech.edu%
} for optical identifications. We also compiled broad-band radio spectra
and collected VLBI images of the sources made at lower frequencies.

The optical identifications yielded 30 quasars, 5 galaxies and 2 unclassified
sources. One of the sources (PMN~J0527$-$1241) turned out to be
a planetary nebula. Indeed, this source remained undetected with the
Japanese VLBI Exploration of Radio Astrometry (VERA) array at 22~GHz
\citep{Petrov}, consistently with an upper limit of 0.11 Jy for the
correlated flux density on the baselines of VERA. This source is therefore
Galactic and its radio structure is resolved, so we excluded it from
our final list. % which consists of 37 radio sources (Table~\ref{sources}). 
We give the name, the 86 GHz flux density interpolated from
WMAP measurements, the spectral index calculated from WMAP flux densities,
the redshift, and the optical identification of 37 extragalactic objects
selected as potential new mm-VLBI targets in Table~\ref{sources}.

VLBI images at lower frequencies (2.3 and 8.4~GHz) were collected
from the Very Long Baseline Array (VLBA) calibrator list%
\footnote{http://www.vlba.nrao.edu/astro/calib/index.shtml%
} maintaned by the U.S. National Radio Astronomy Observatory (NRAO).
Maps were found here for all but one of our 37 sources. However, the
missing object, PKS J1332$+$0200 was imaged with the VLBA at 5~GHz
\citep{Fomal}. This proves that our remaining 37 objects are indeed
compact extragalaxtic sources.

For compiling the broad-band radio spectra, our primary source of
information was a catalogue made with the Russian RATAN-600 radio
telescope \citep{Kovalev}. It is based on instantaneous flux density
measurements at six different frequencies between 1 and 22~GHz. The
effect of time variability is excluded due to the simultaneous measurements.
The \citet{Kovalev} catalogue contains sources with declination between
$-30^{\circ}<\delta <$ $+43^{\circ}$. We found complete 6-frequency data
for 20 of our sources. Another 6 sources were partially covered in
the frequency range. For the remaining sources, we again used the
NED to search for 1.4 and 4.8 GHz flux densities from the literature.
Fig.~\ref{spectrum} and Fig.~\ref{spectrumNED} show broad-band spectra for the 26
sources found in \citet{Kovalev}, with the WMAP flux densities at
41, 61 and 94~GHz included. The majority of our sources have flat
radio spectra over nearly two orders of magnitude in frequency. However,
a number of sources can be classified as High-Frequency Peakers (HFP)
that show convex spectrum peaking at above $\sim$5 GHz, and therefore
negative spectral index in the WMAP bands (e.g. PMN J0457$-$2324,
GB6 J0825$+$0309, PMN J2000$-$1476, PMN J2131$-$1207; Fig.~\ref{spectrum}).
When imaged with VLBI, these sources are expected to be generally
less compact than those with flat spectra in the broadest range of
frequencies, but more compact than the Gigahertz-Peaked Spectrum (GPS)
sources of which the spectrum peaks at lower frequencies \citep[e.g.][]{Vollmer}.

We also looked for supplementary broad-band radio spectral
information in the SPECFIND catalogue \citep{Vollmer2010} which was
compiled from a large literature database of non-simultaneous flux
density measurements. For the available sources, the broad-band spectral
index is also listed in Table~\ref{sources} for comparison. 
Note that there is indication of a spectral break in some cases, 
thus the linear spectral fits are not always the best representations
of the broad-band spectrum.

We found indications of variability in several cases, because the
low- and high-band measurements were made at different epochs. A
simple visual inspection of the spectral shapes in Fig.~\ref{spectrum}
leads to a conservative estimate that at least $\sim$25\% of our
sources are variable on the level of $\sim$50\% or more. Particular
examples are PMN J0132$-$1654, PMN J0137$-$2430, PMN J0808$-$0751,
PMN J1037$-$2934, PMN J1337$-$1257, GB6 J1635$+$3808, GB6 J1753$+$2847, and PMN J2229$-$0832. Both the flat-spectrum and the variability are indicative of the compact radio
structure.

%%%%%%%%%%%%%%%%%%%% Table 1
%
\begin{table}
\caption{The list of the 37 bright WMAP extragalactic point sources selected
as potential new mm-VLBI targets}

\label{sources} \begin{tabular}{clccccc}
\hline 
WMAP name  & Other name & $S_{86}$ {[}Jy{]}  & $\alpha_{WMAP}$  & $\alpha_{V}$ & $z$  & ID \tabularnewline
\hline 
J0012$-$3952  & PMN J0013$-$3954 &  1.1$\pm$0.4  & 0.53        & ... & ...  & U \tabularnewline
J0132$-$1653  & PMN J0132$-$1654 & 1.1$\pm$0.4  & 0.37         & $-$0.22 & 1.02  & Q \tabularnewline
J0137$-$2428  & PMN J0137$-$2430 & 1.2$\pm$0.4  & $-$0.52      & $-$0.40 & 0.83  & Q \tabularnewline
J0152+2208  & GB6 J0152+2206 & 1.2$\pm$0.4  & 0.10           & $-$0.12 & 1.32  & Q \tabularnewline
J0403$-$3604  & PMN J0403$-$3605 & 3.3$\pm$0.3  & $-$0.03      & 0.28 & 1.41  & Q \tabularnewline
J0428$-$3757  & PMN J0428$-$3756 & 1.4$\pm$0.4  & 0.29         & 0.10 & 1.11  & Q \tabularnewline
J0453$-$2806  & PMN J0453$-$2807 & 1.0$\pm$0.4  & 0.00         & $-$0.12 & 2.55  & Q \tabularnewline
J0456$-$2322  & PMN J0457$-$2324 & 1.8$\pm$0.4  & $-$0.45      & 0.12 & 1.00  & Q \tabularnewline
J0757+0957  & GB6 J0757+0956 & 1.4$\pm$0.5  & $-$0.15        & 0.08 & 0.26  & Q \tabularnewline
J0808$-$0750  & PMN J0808$-$0751 & 1.1$\pm$0.4  & 0.11         & $-$0.31 & 1.83  & Q \tabularnewline
J0825+0311  & GB6 J0825+0309 & 1.2$\pm$0.5  & $-$0.34        & 0.14 & 0.50  & Q \tabularnewline
J0840+1312  & GB6 J0840+1312 & 1.1$\pm$0.5  & $-$0.54        & $-$0.75 & 0.68  & Q \tabularnewline
J0914+0248  & GB6 J0914+0245 & 1.1$\pm$0.5  & 0.25           & $-$0.04 & 0.42  & G \tabularnewline
J1037$-$2934  & PMN J1037$-$2934 & 1.9$\pm$0.4  & 0.43         & 0.19 & 0.31  & Q \tabularnewline
J1127$-$1858  & PMN J1127$-$1857 & 1.2$\pm$0.4  & 0.11         & 0.21 & 1.05  & Q \tabularnewline
J1219+0549  & 1Jy 1216+06 & 1.3$\pm$0.4  & $-$0.54           & ... & 0.01  & G \tabularnewline
J1222+0414  & GB6 J1222+0413 & 1.0$\pm$0.4  & 0.01           & $-$0.06 & 0.96  & Q \tabularnewline
J1246$-$2547  & PMN J1246$-$2547 & 1.2$\pm$0.4  & $-$0.45      & 0.04 & 0.63  & Q \tabularnewline
J1258$-$3158  & PMN J1257$-$3154 & 1.2$\pm$0.4  & 0.13         & 0.03 & 1.92  & Q \tabularnewline
J1316$-$3337  & PMN J1316$-$3339 & 1.4$\pm$0.4  & $-$0.30      & $-$0.07 & 1.21  & Q \tabularnewline
J1332+0200  & GB6 J1332+0200 & 1.2$\pm$0.4  & 0.35           & $-$0.41   & 0.21  & G \tabularnewline
J1337$-$1257  & PMN J1337$-$1257 & 4.8$\pm$0.5  & $-$0.29      & $-$0.18 & 0.53  & Q \tabularnewline
J1356+1919  & GB6 J1357+1919 & 1.1$\pm$0.4  & $-$0.20        & $-$0.53 & 0.72  & Q \tabularnewline
J1512$-$0904  & 1Jy 1510$-$08 & 1.7$\pm$0.4  & $-$0.13         & $-$0.19 	 & 0.36  & Q \tabularnewline
J1516+0014  & GB6 J1516+0015 & 1.0$\pm$0.4  & $-$0.12        & $-$0.56 & 0.05  & G \tabularnewline
J1517$-$2421  & PMN J1517$-$2422 & 1.8$\pm$0.5  & $-$0.18      & 0.05 & 0.04  & G \tabularnewline
J1635+3807  & GB6 J1635+3808 & 3.4$\pm$0.4  & $-$0.03        & $-$0.01 & 1.81  & Q \tabularnewline
J1734+3857  & GB6 J1734+3857 & 1.0$\pm$0.4  & $-$0.11        & 0.27 & 0.97  & Q \tabularnewline
J1753+2848  & GB6 J1753+2847 & 1.1$\pm$0.4  & $-$0.78        & 0.09 & 1.11  & U \tabularnewline
J1849+6705  & GB6 J1849+6705 & 1.5$\pm$0.3  & 0.08           & $-$0.20 & 0.65  & Q \tabularnewline
J1923$-$2105  & PMN J1923$-$2104 & 2.2$\pm$0.5 &  $-$0.10      & ... & 0.87  & Q \tabularnewline
J1958$-$3845  & PMN J1957$-$3845 & 1.7$\pm$0.5  & $-$0.76      & 0.20 & 0.63  & Q \tabularnewline
J2000$-$1749  & PMN J2000$-$1748 & 1.4$\pm$0.4  & $-$0.09      & 0.42 & 0.65  & Q \tabularnewline
J2005+7755  & 1Jy 2007+77 & 1.1$\pm$0.4  & 0.41              & 0.30 & 0.34  & Q \tabularnewline
J2131$-$1207  & PMN J2131$-$1207 & 1.4$\pm$0.4  & $-$0.62      & 0.09 & 0.50  & Q \tabularnewline
J2229$-$0833  & PMN J2229$-$0832 & 1.7$\pm$0.5  & $-$0.26      & 0.14 & 1.55  & Q \tabularnewline
J2354+4550  & GB6 J2354+4553 & 1.3$\pm$0.4  & 0.32           & $-$0.25 & 1.99  & Q \tabularnewline
\hline 
\end{tabular}\\
{\small {Notes: Col. 1 -- WMAP source name; Col. 2 -- Other name; Col. 3 -- estimated
86-GHz flux density (Jy); Col. 4 -- spectral index calculated from
WMAP flux densities \citep{Chen}; Col. 5 -- spectral index from the \citet{Vollmer2010} catalogue; Col. 6 -- redshift from NED; Col.
7 -- optical identification from NED (Q=quasar, G=galaxy, U=unclassified)} }
\end{table}

%%%%%%%%%%%%%%%%%%%%

%%%%%%%%%%%%%%%%%%%%%%%%%%%%%%%%% visszacserelni egy spktrumra
%%%%%%%%%%%%%%%%%%%% Fig. 2
%
\begin{figure}
\begin{center}
\includegraphics[width=6cm]{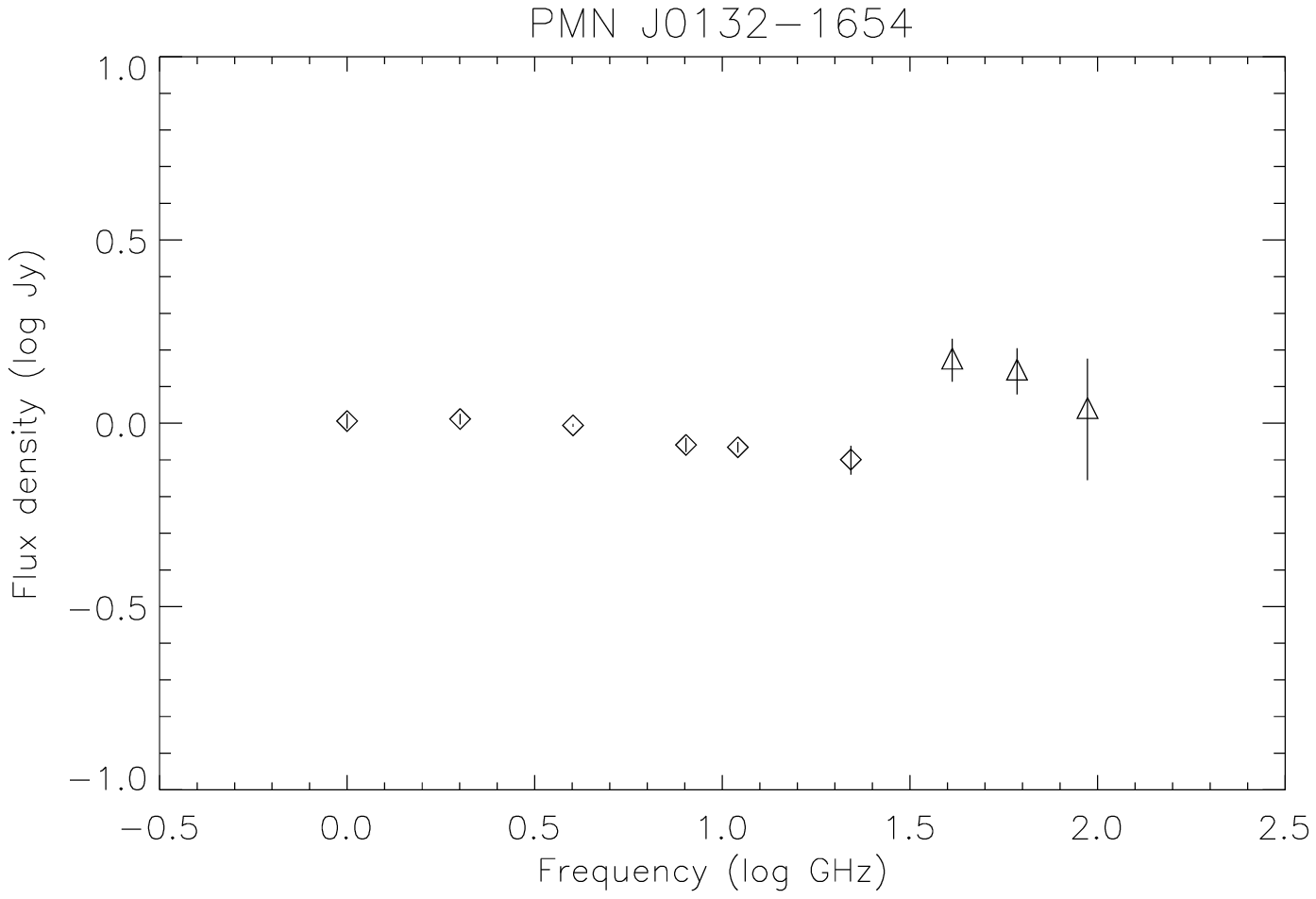}
\includegraphics[width=6cm]{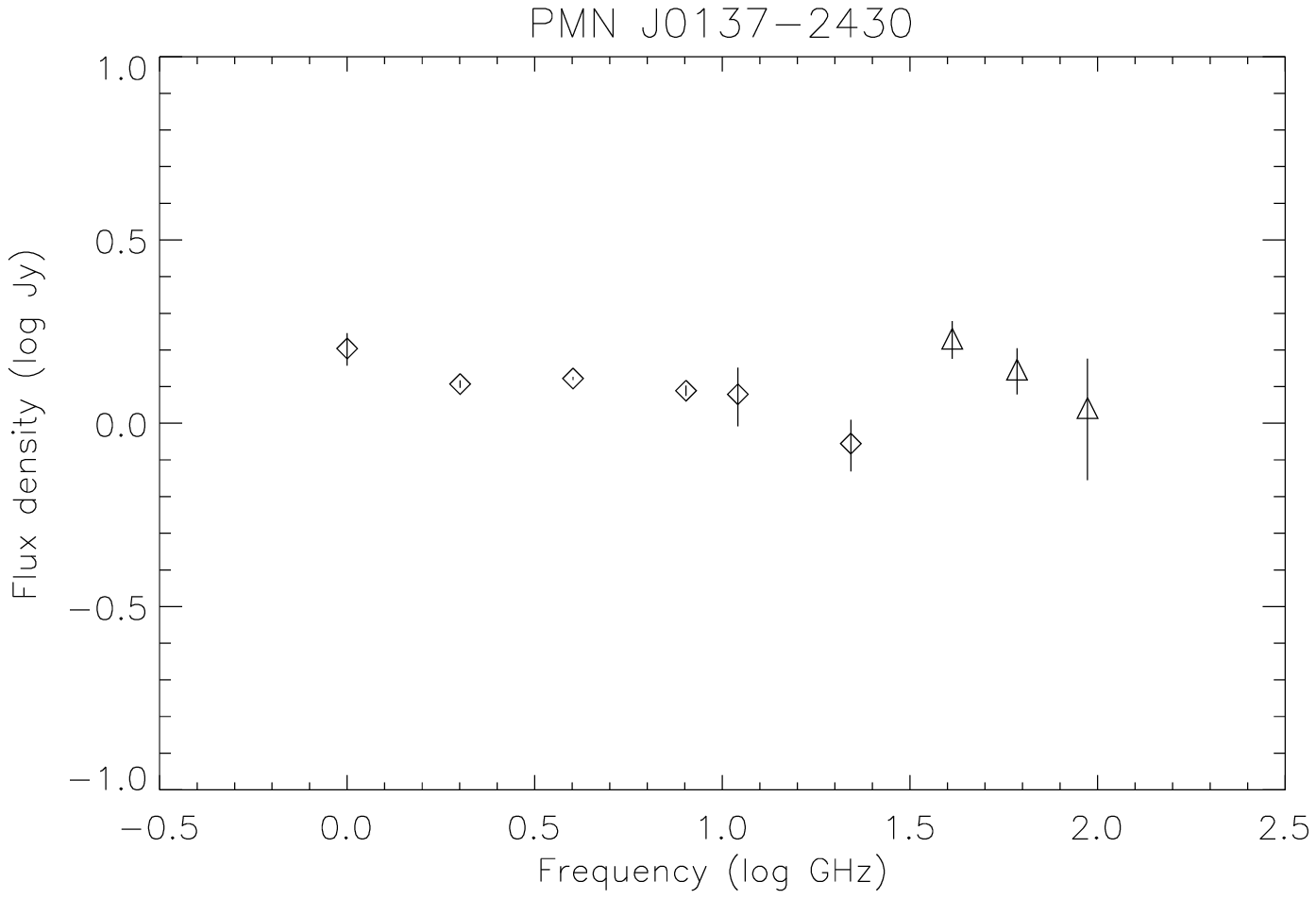}
\includegraphics[width=6cm]{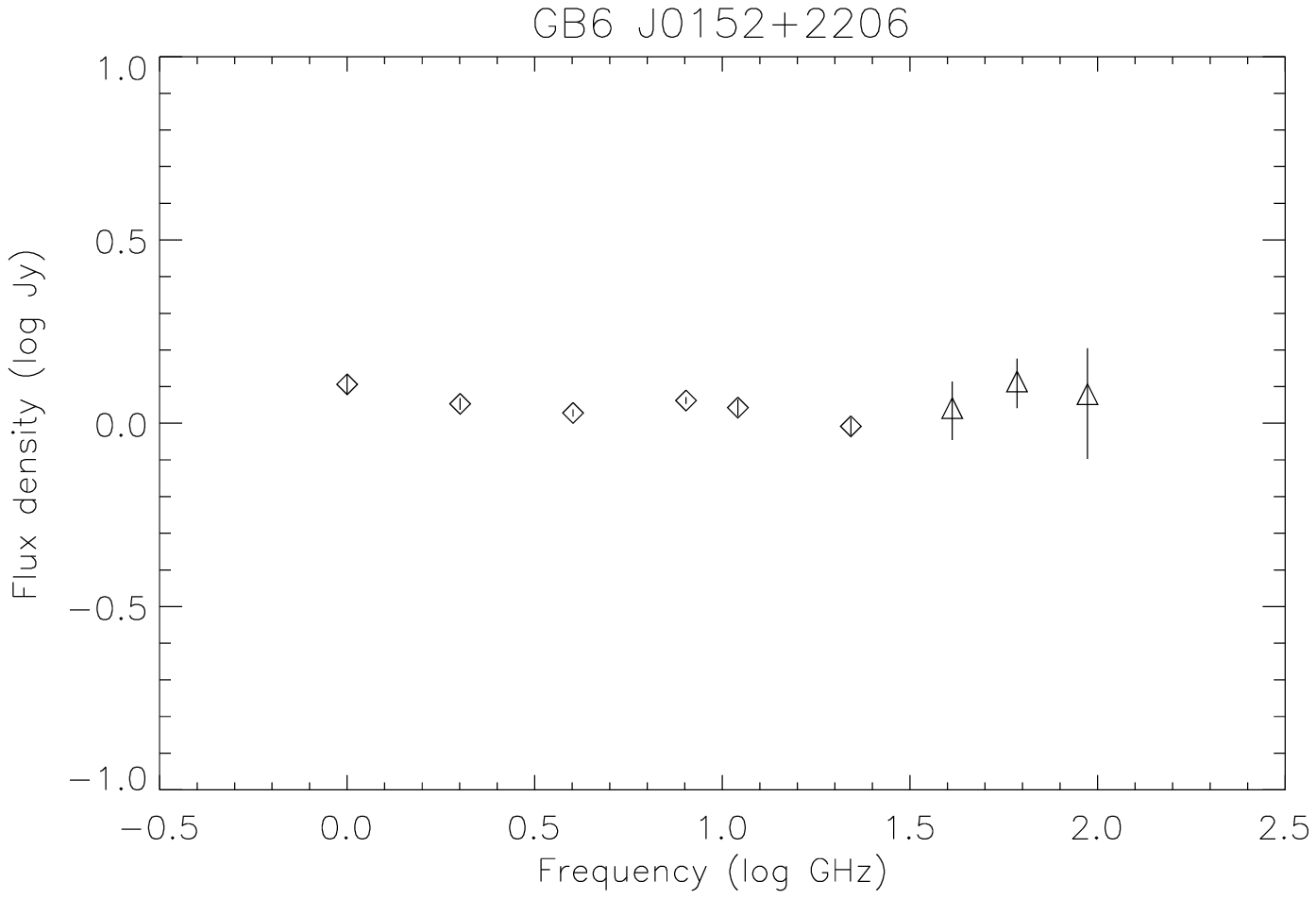}
\includegraphics[width=6cm]{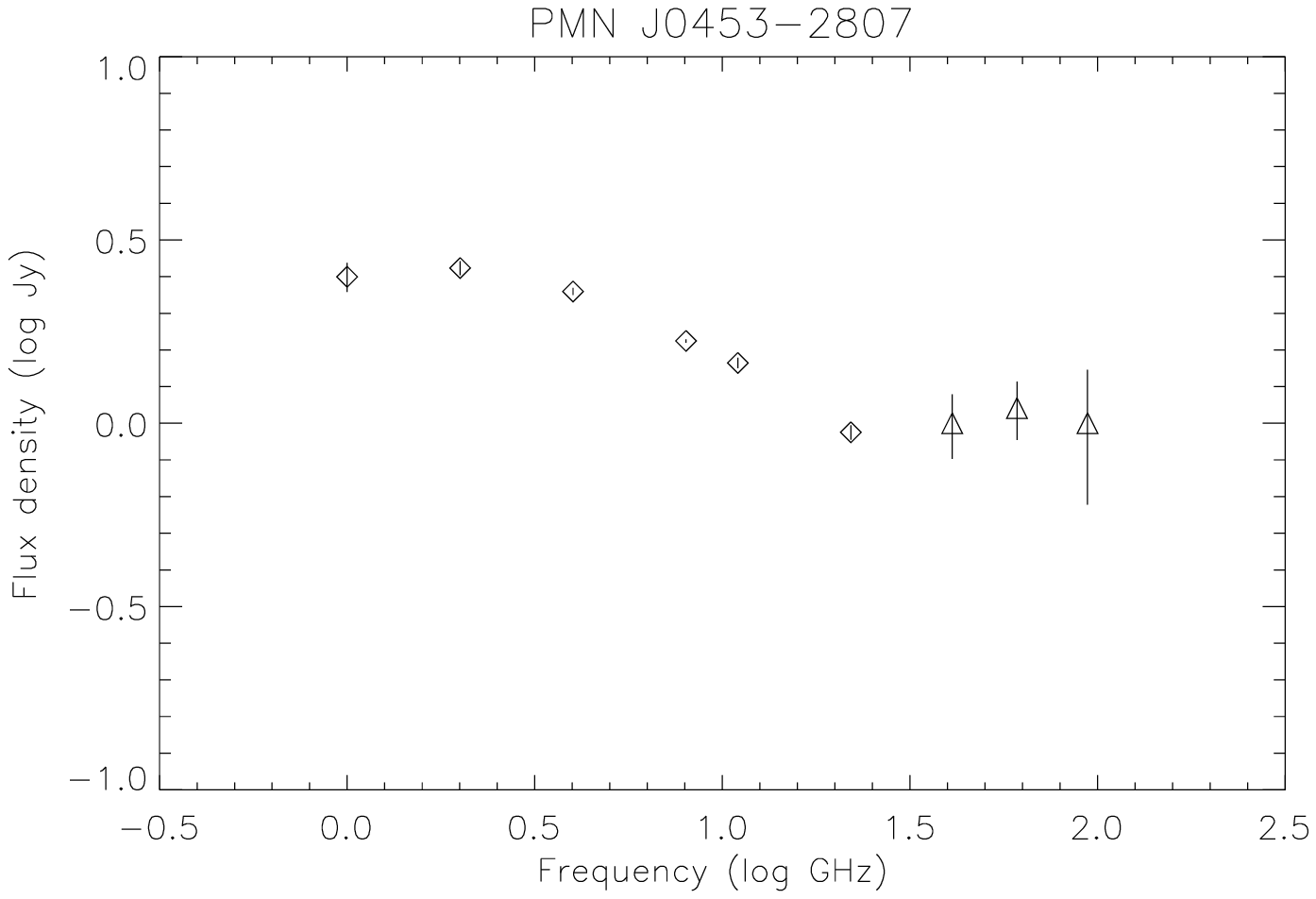}
\includegraphics[width=6cm]{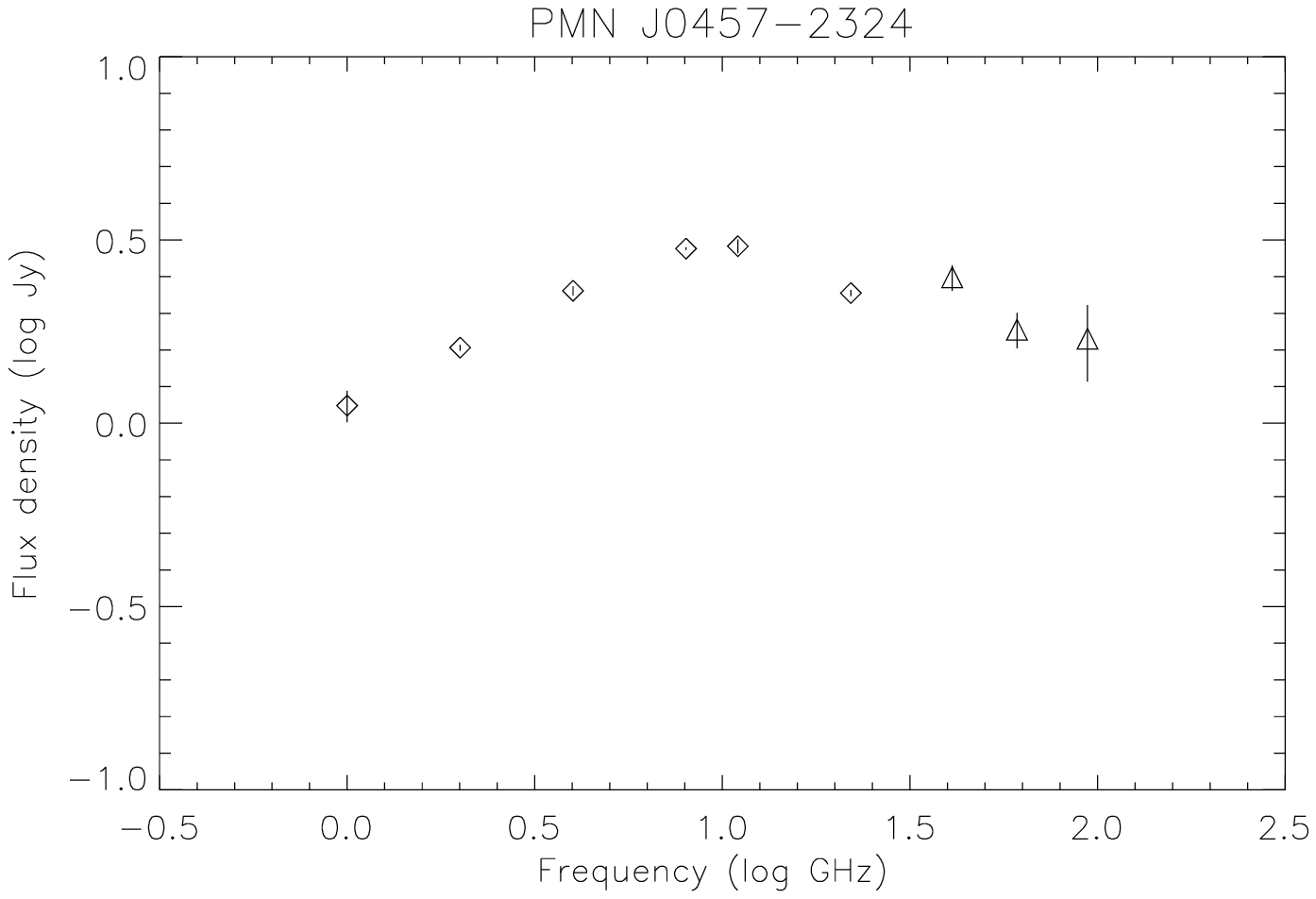}
\includegraphics[width=6cm]{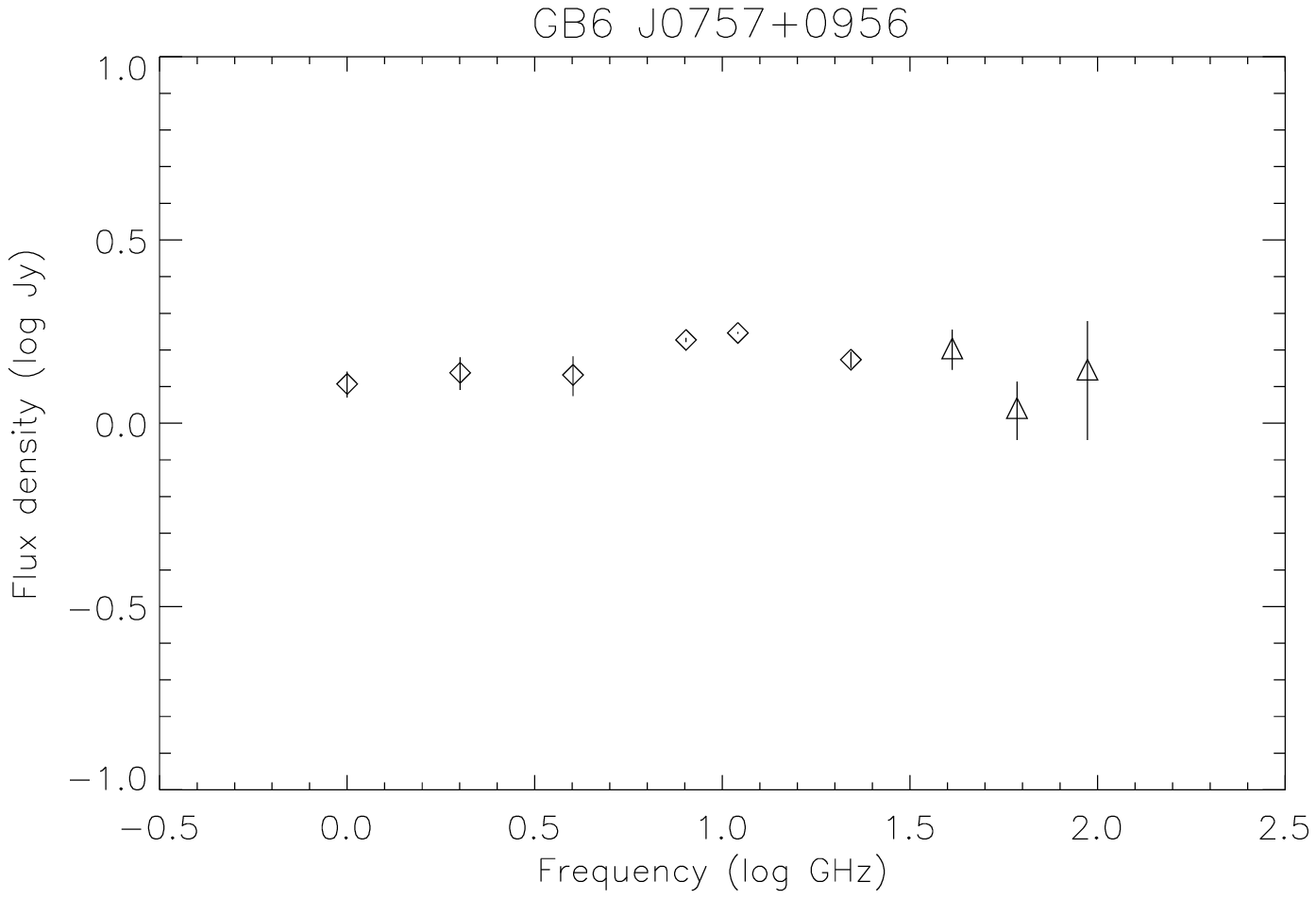}
\includegraphics[width=6cm]{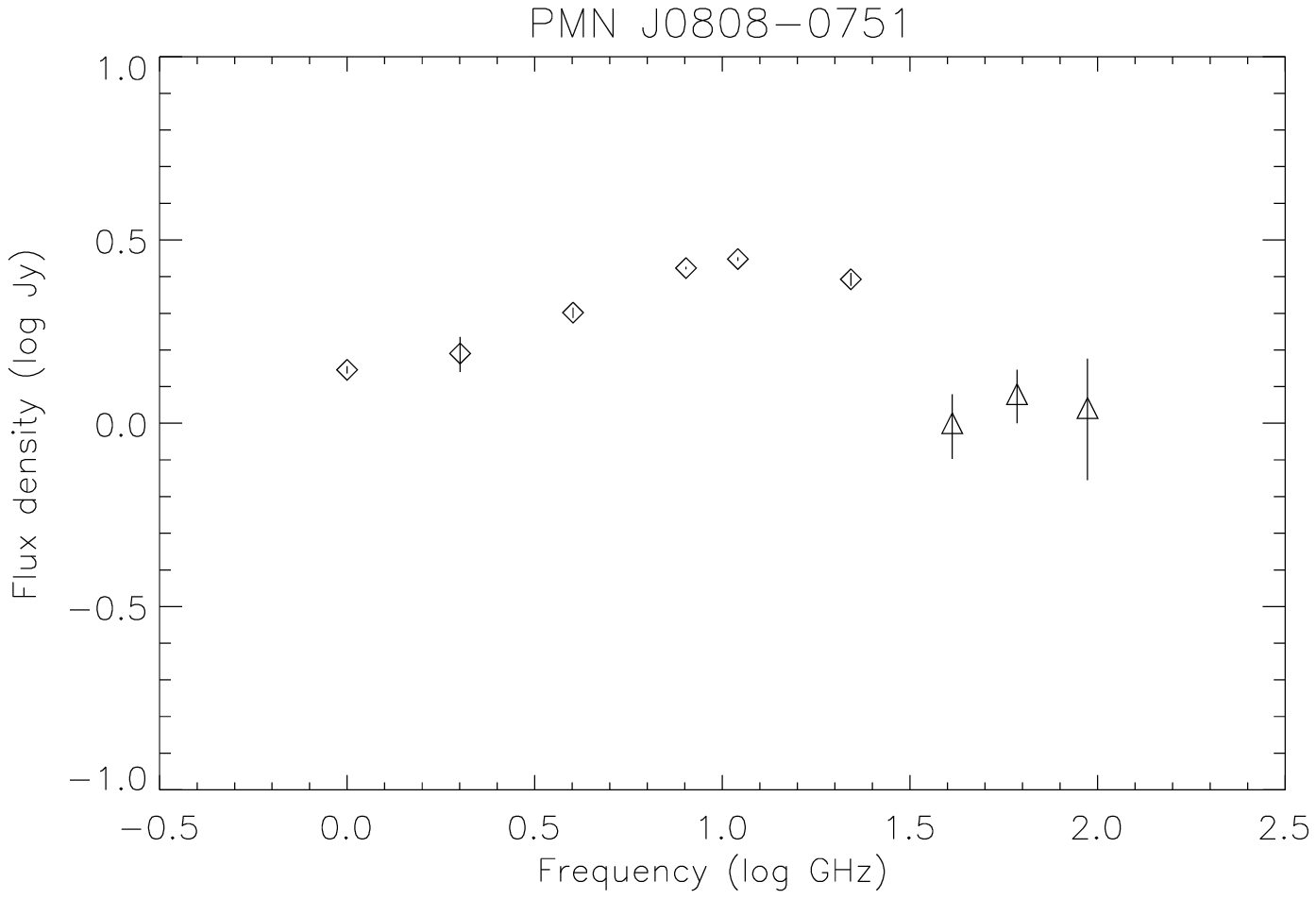}
\includegraphics[width=6cm]{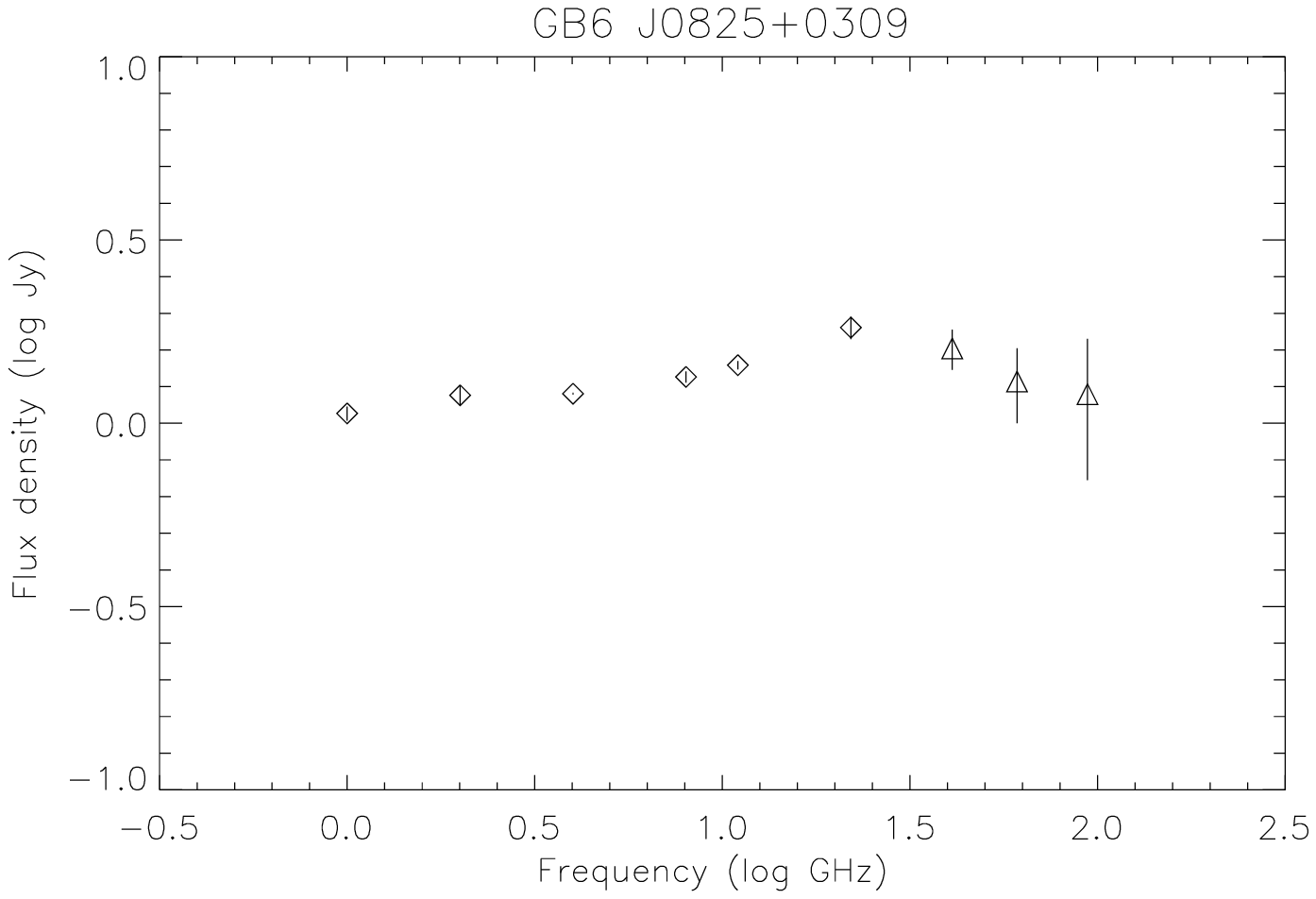}
\end{center}
\caption{Radio spectrum of the sources, including the 6-frequency
data from the \citet{Kovalev} catalogue (diamonds), and
the WMAP flux densities at 41, 61 and 94 GHz (triangles). }
\label{spectrum} 
\end{figure}

 \addtocounter{figure}{-1}
\begin{figure}
\begin{center}
\includegraphics[width=6cm]{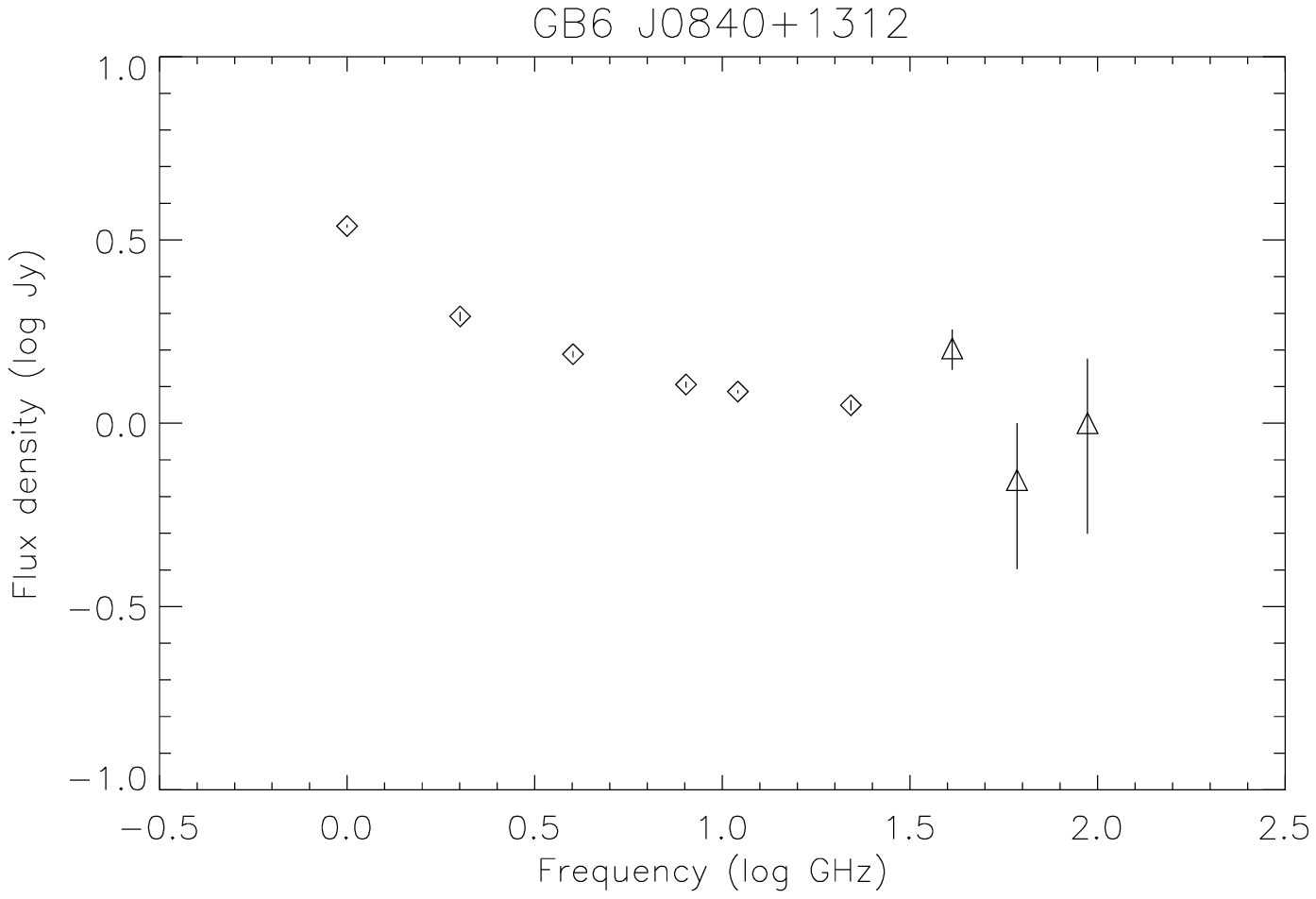}
\includegraphics[width=6cm]{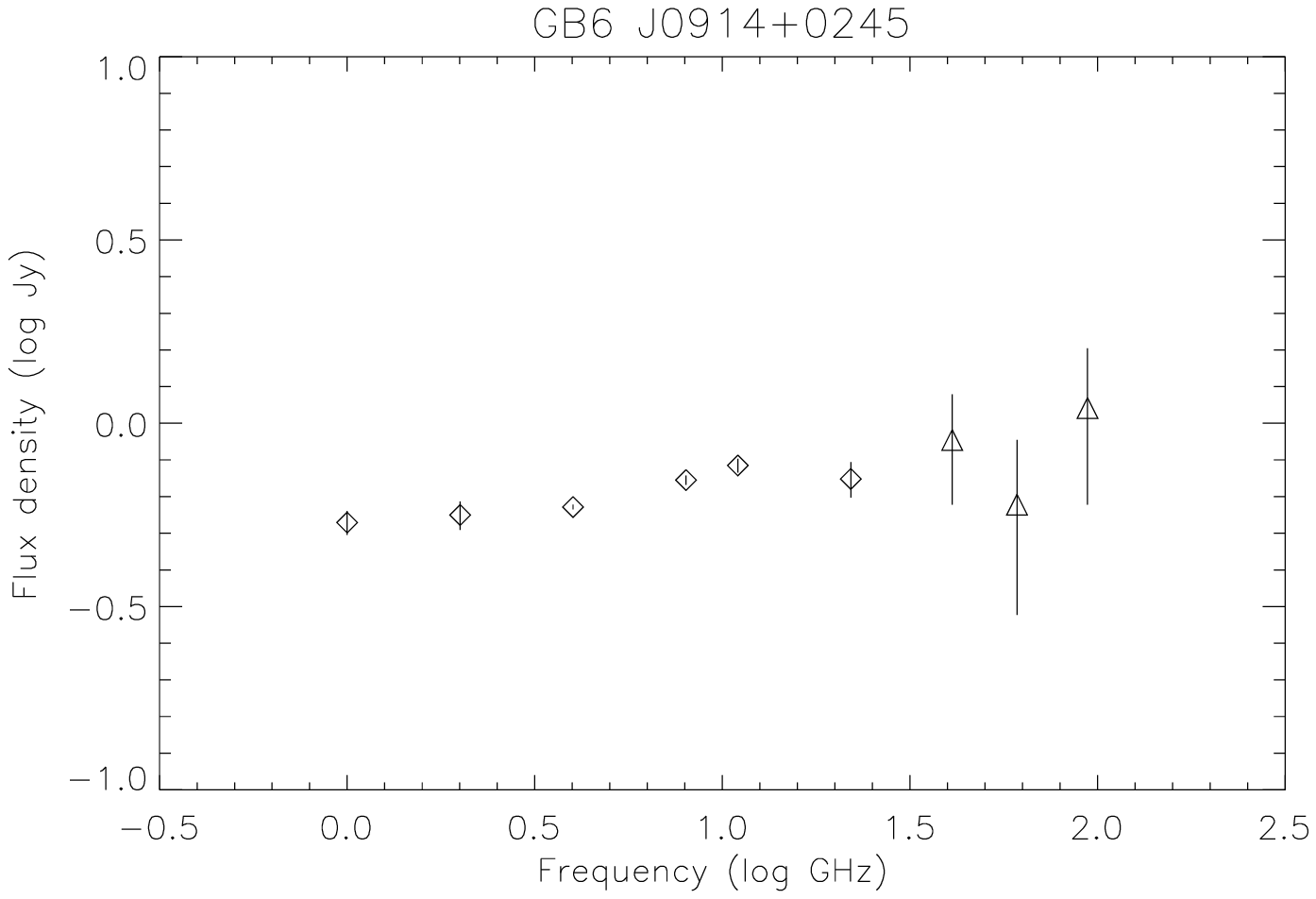}
\includegraphics[width=6cm]{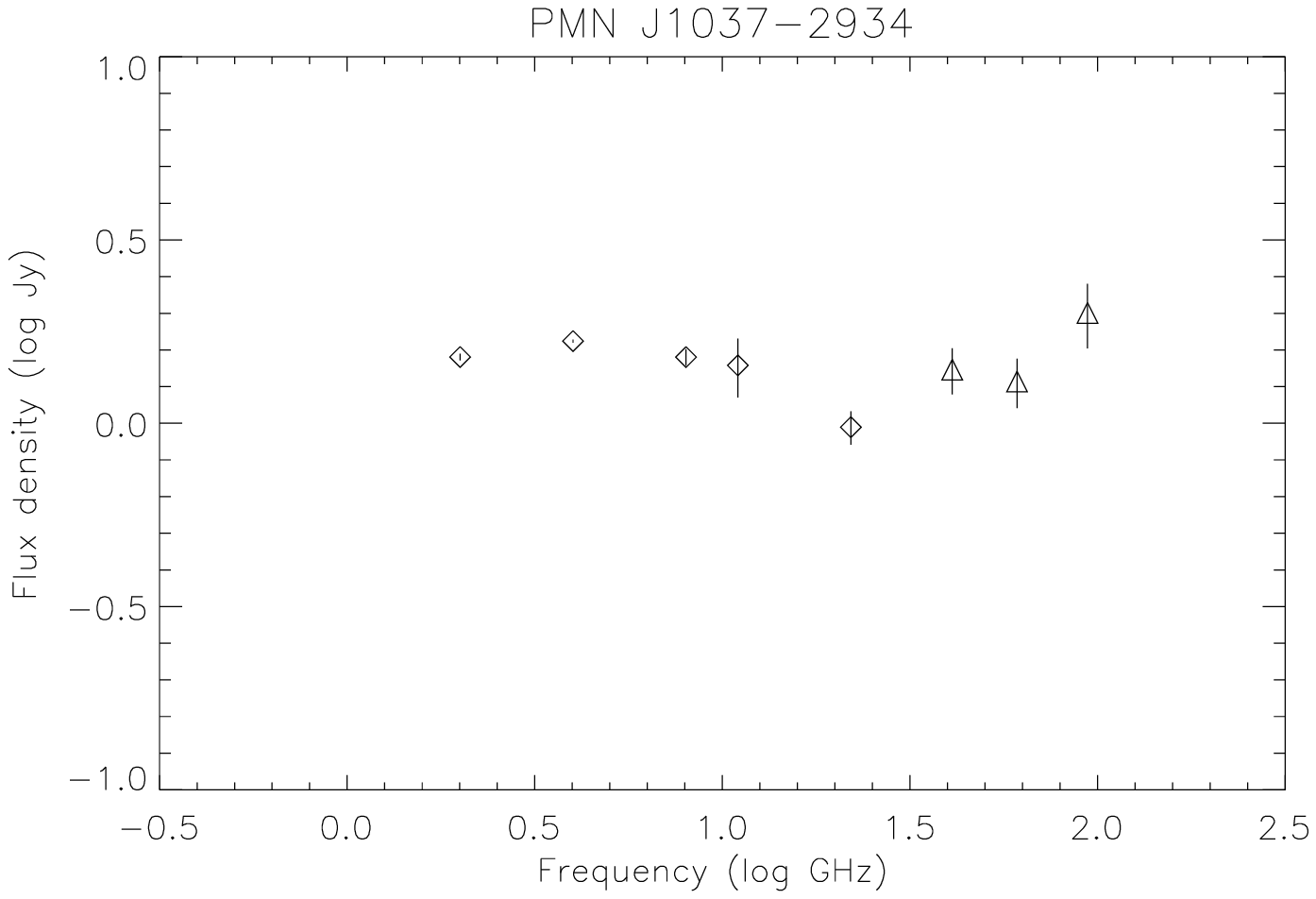}
\includegraphics[width=6cm]{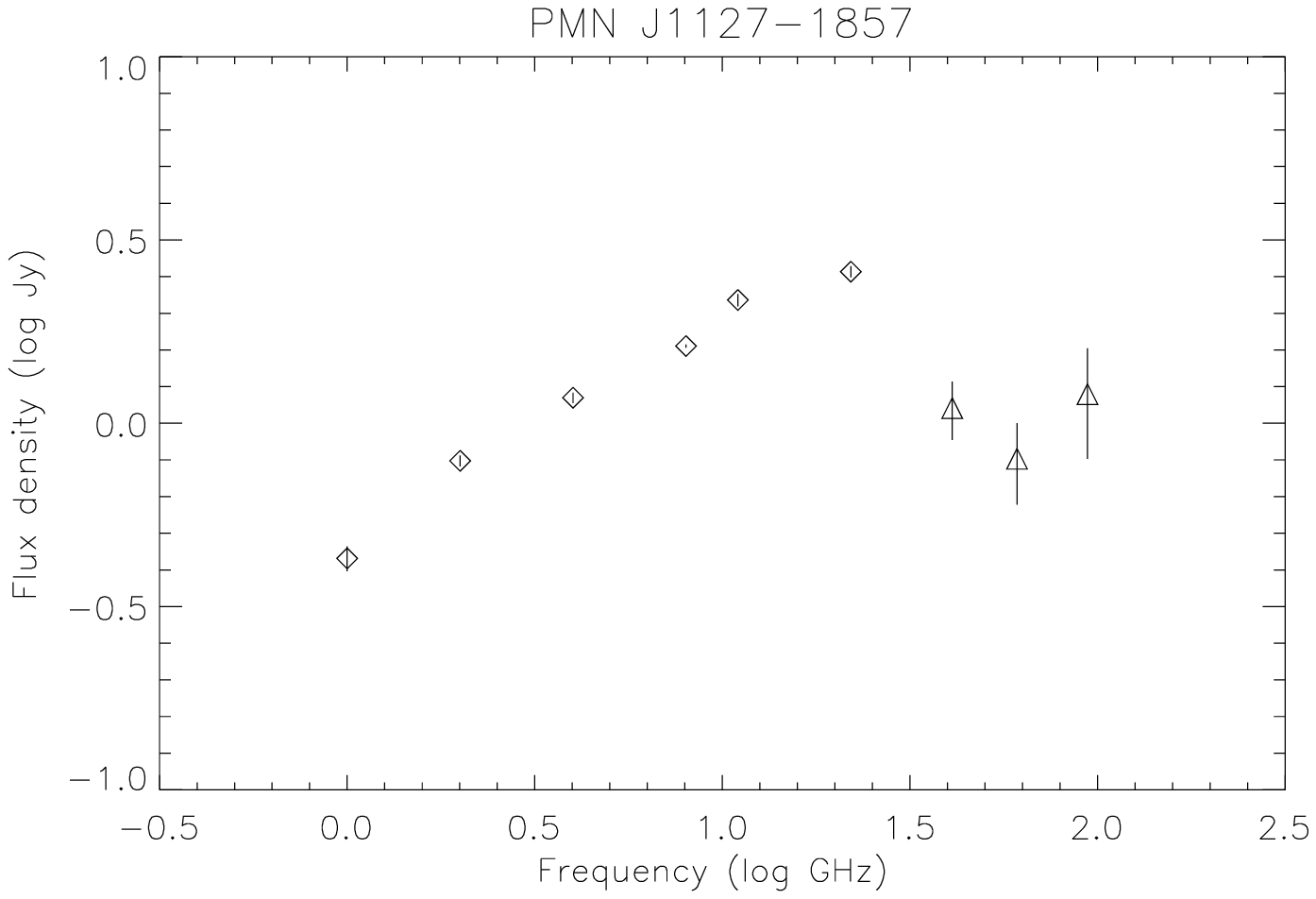}
\includegraphics[width=6cm]{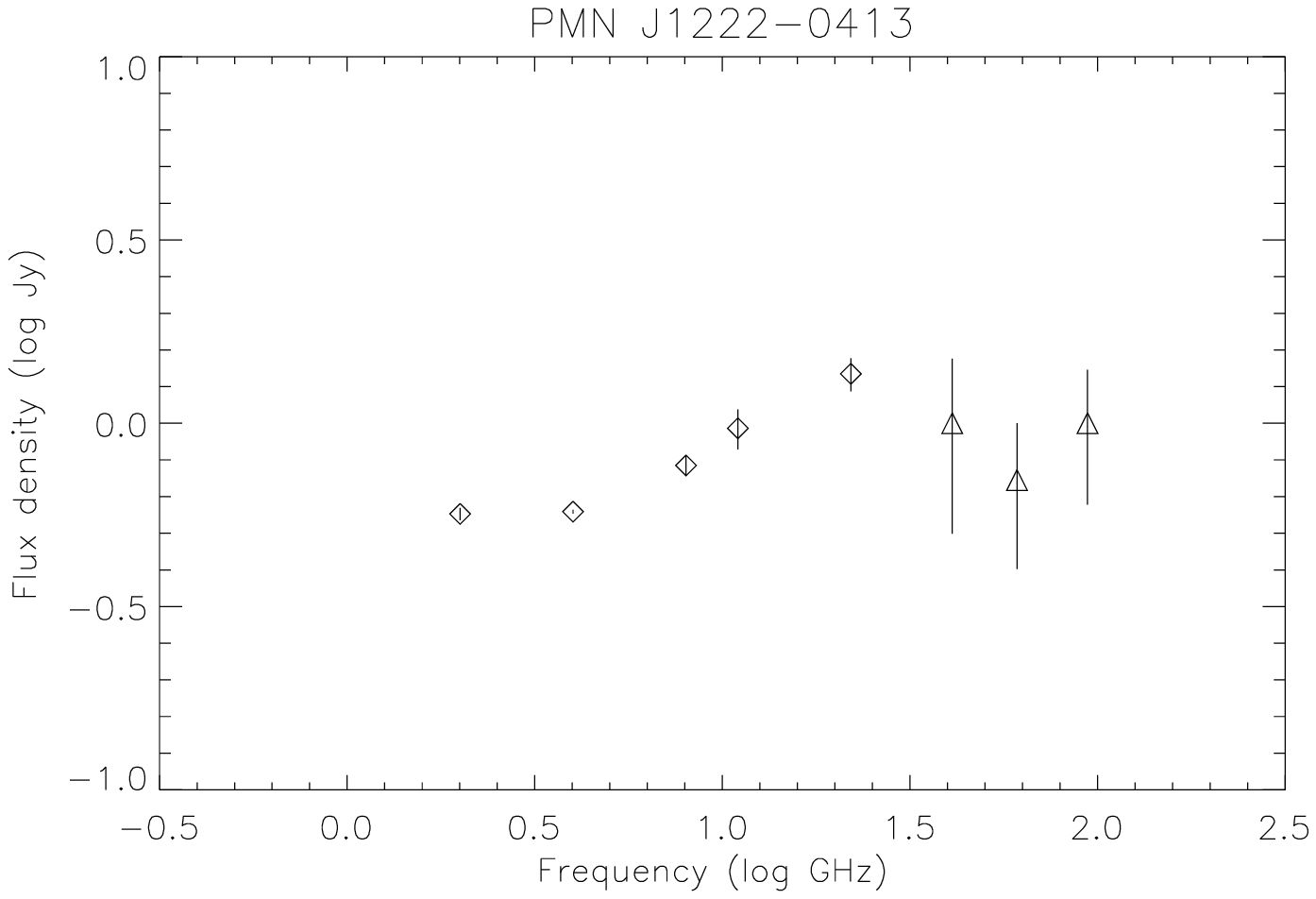}
\includegraphics[width=6cm]{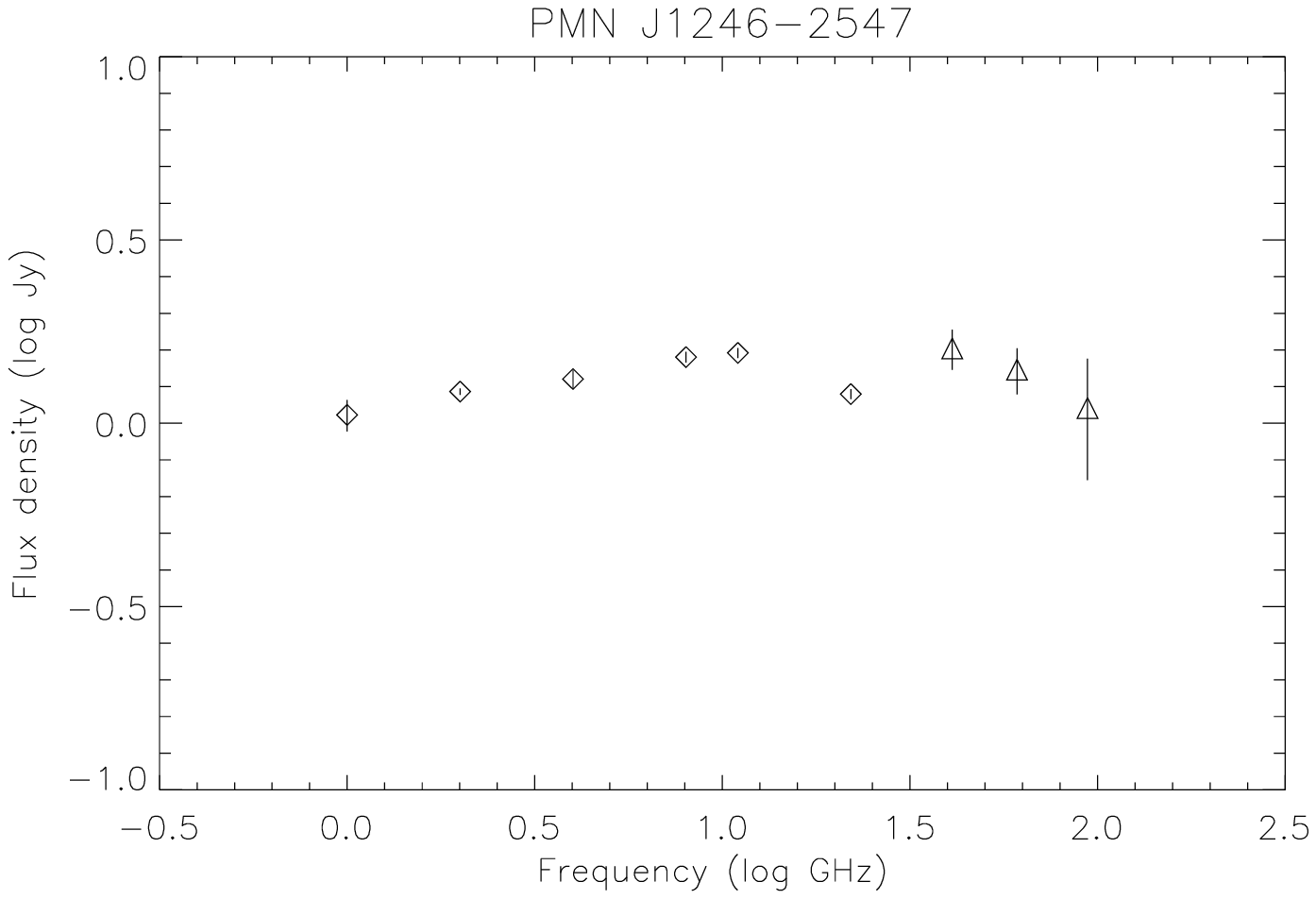}
\includegraphics[width=6cm]{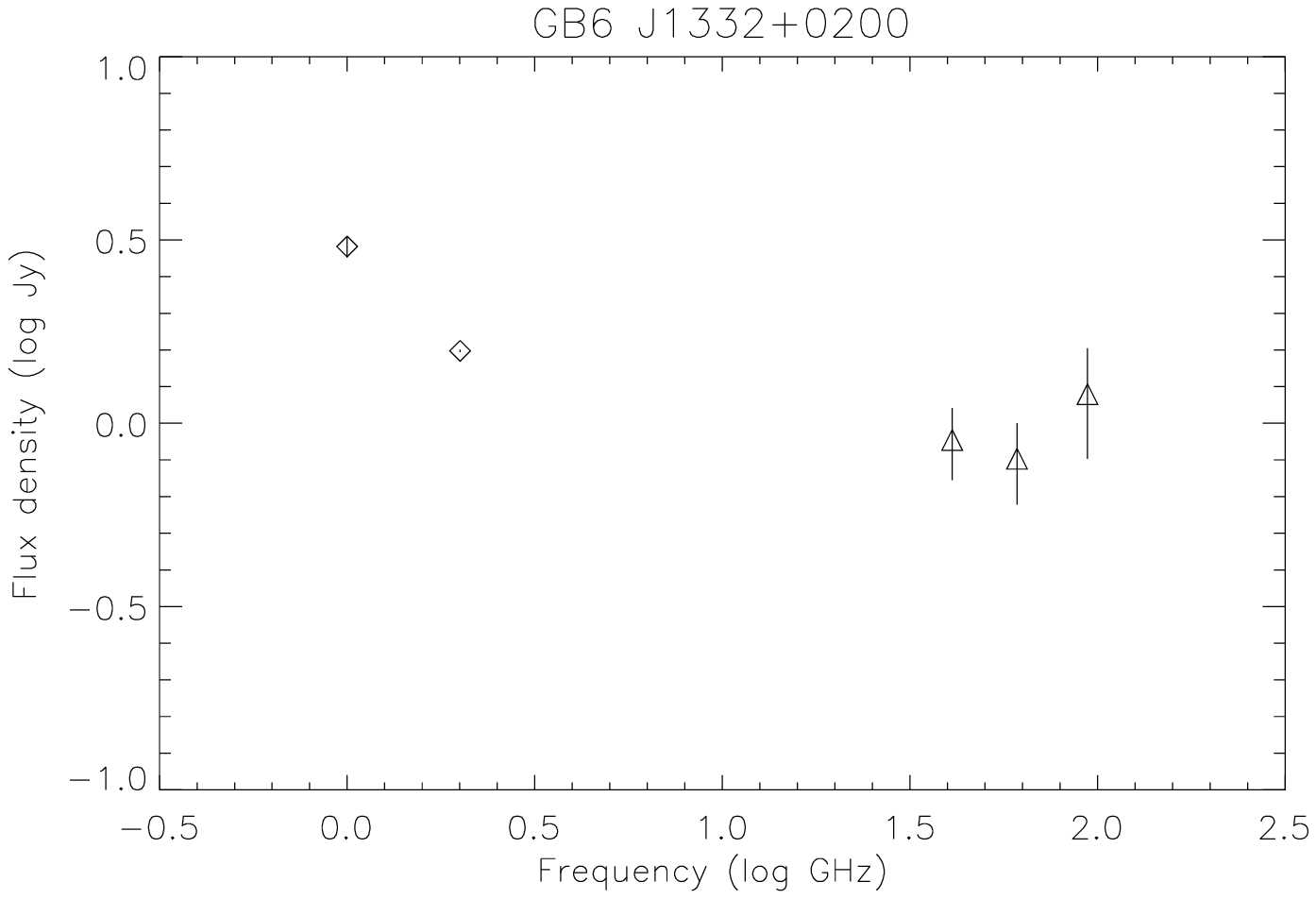}
\includegraphics[width=6cm]{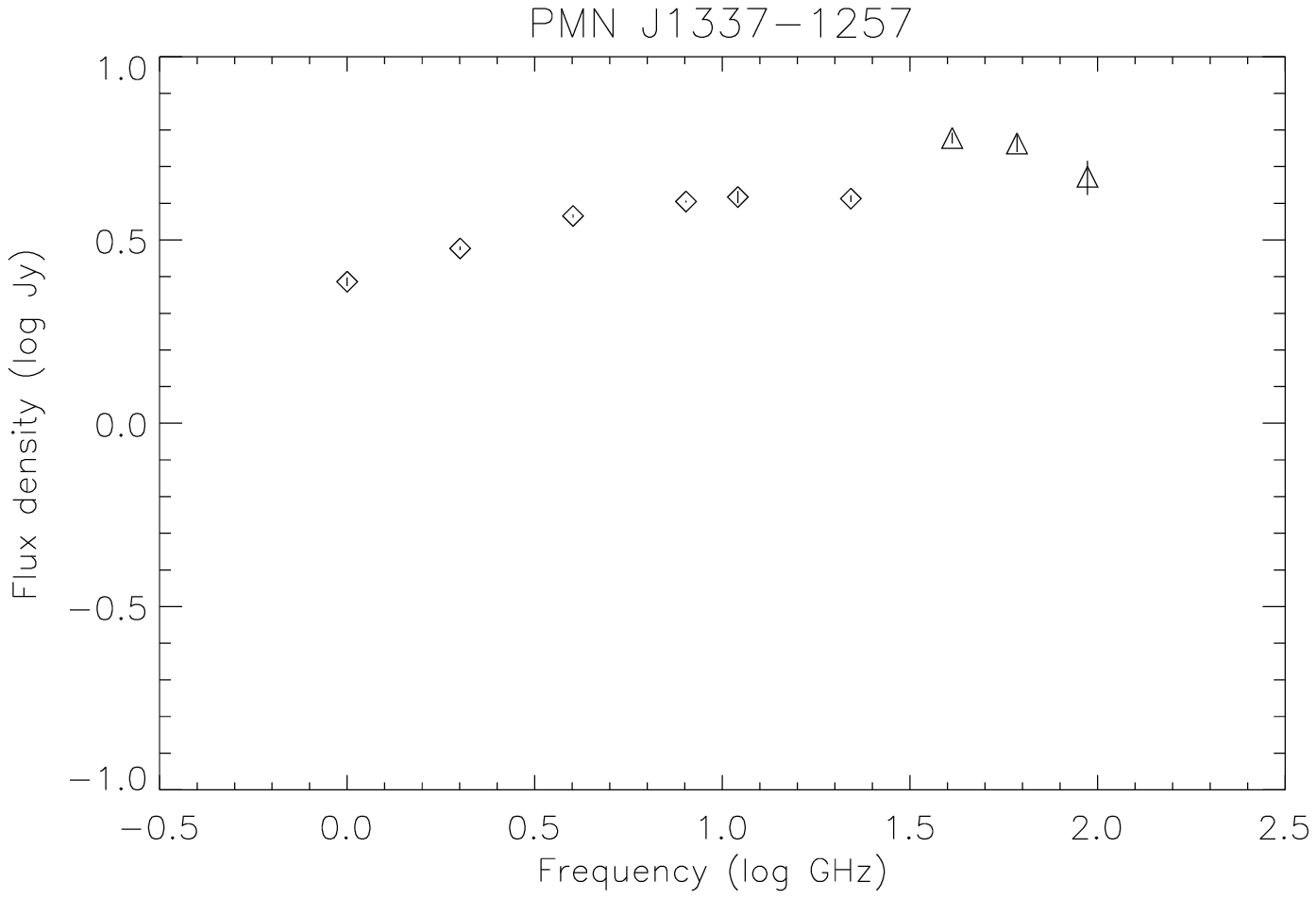}
\end{center}
\caption{ -- Continued }
\label{spectrum} 
\end{figure}
%%%%%%%%%%%%%%%%%%%%

 \addtocounter{figure}{-1}
\begin{figure}
\begin{center}
\includegraphics[width=6cm]{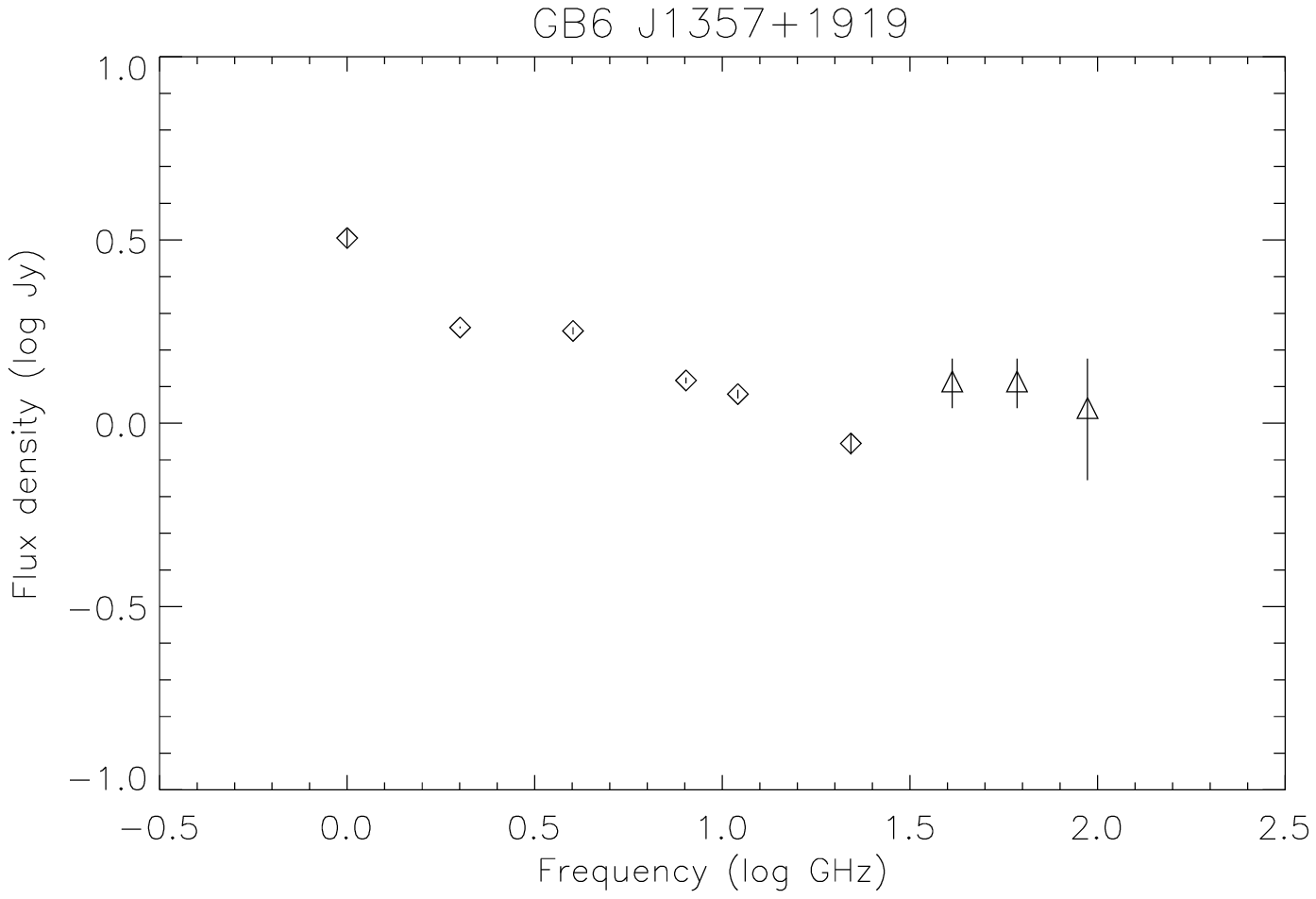}
\includegraphics[width=6cm]{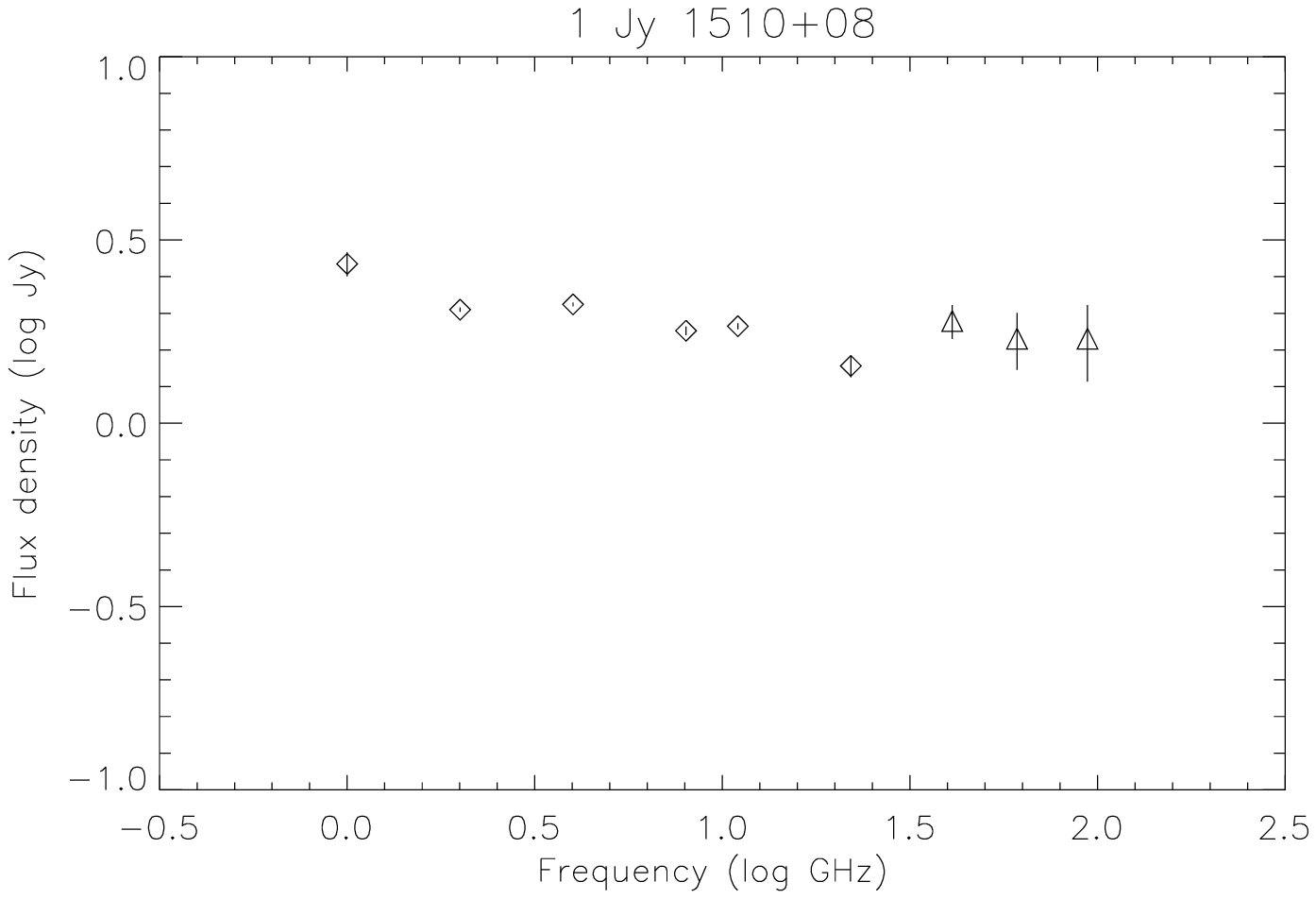}
\includegraphics[width=6cm]{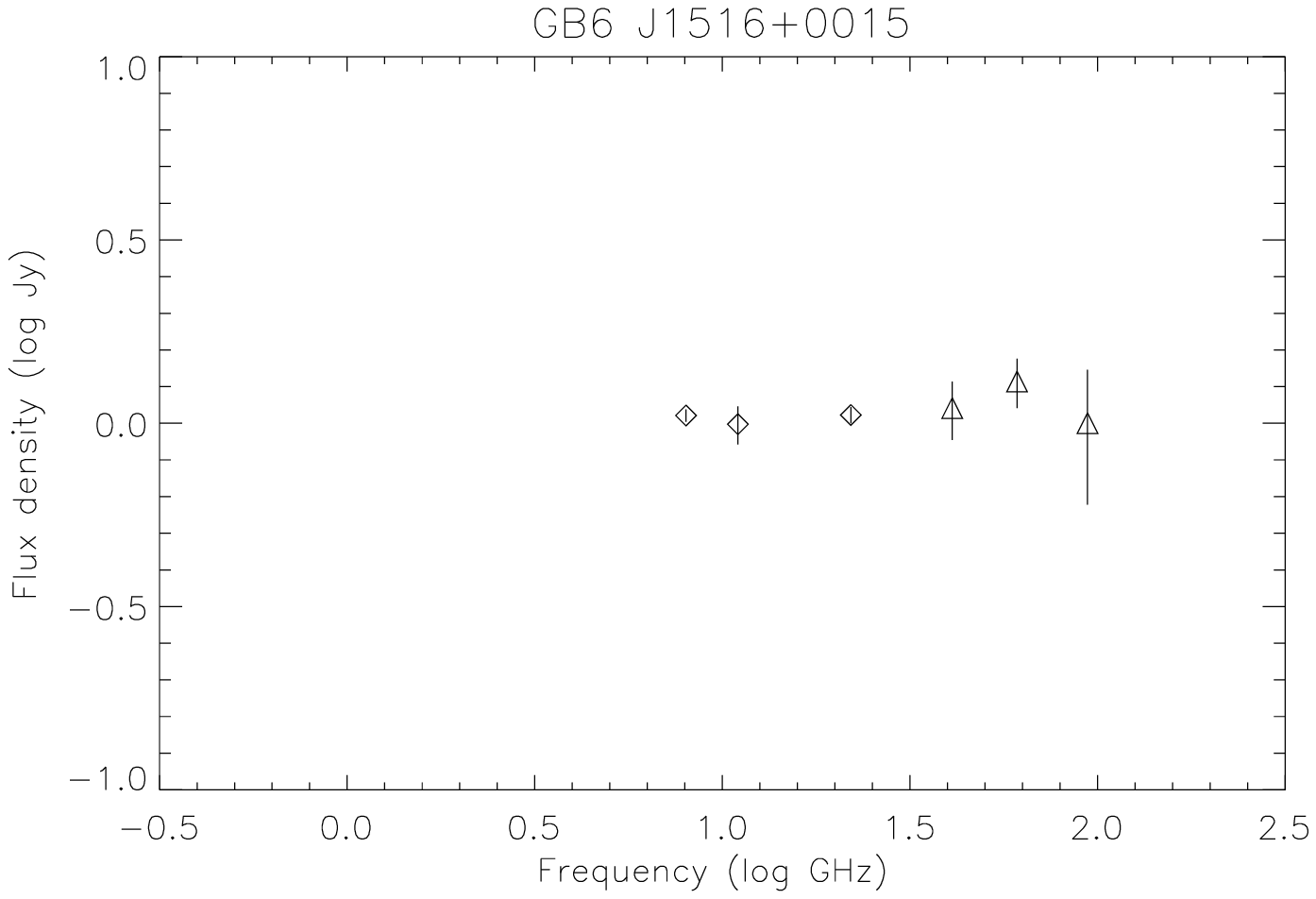}
\includegraphics[width=6cm]{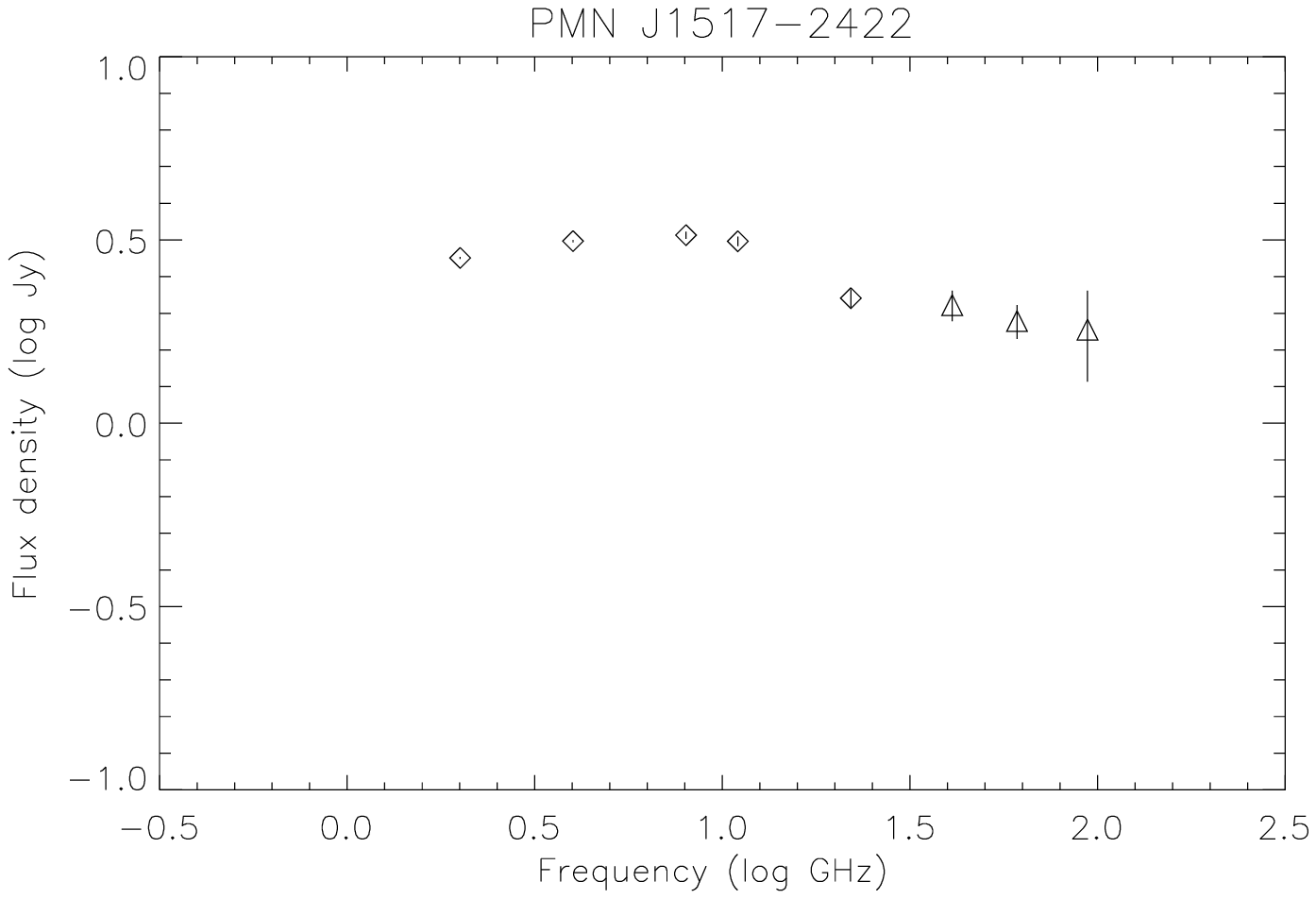}
\includegraphics[width=6cm]{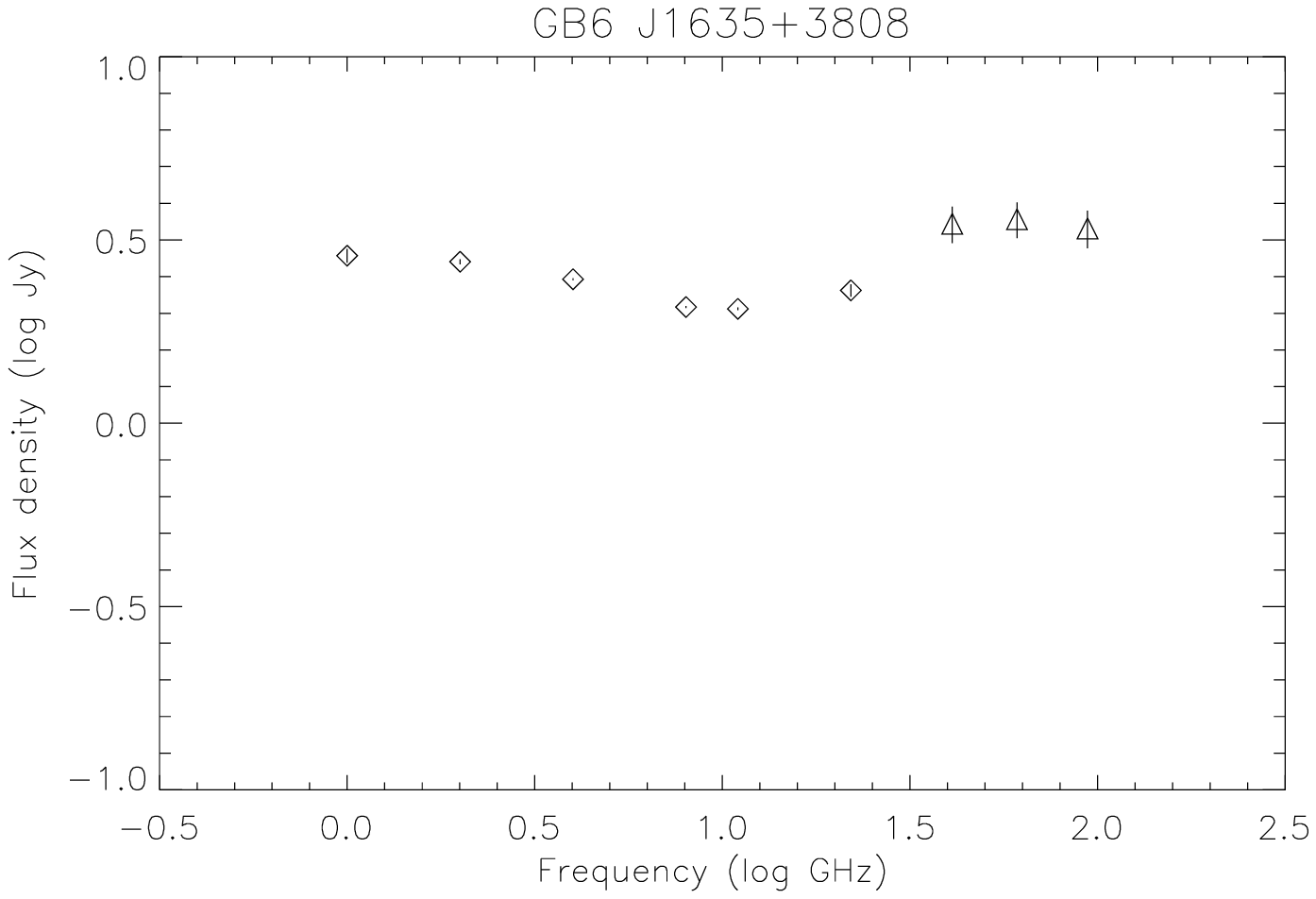}
\includegraphics[width=6cm]{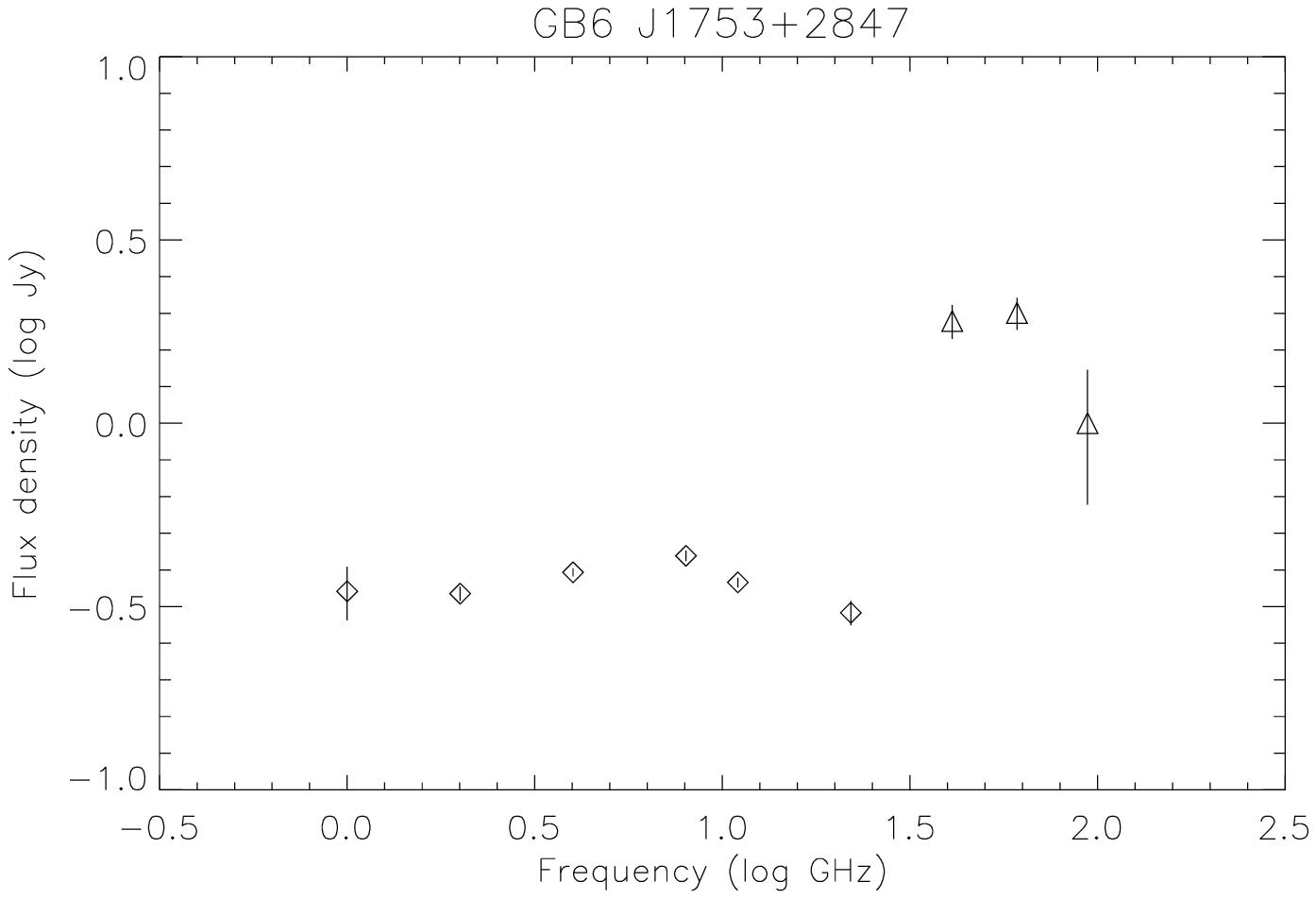}
\includegraphics[width=6cm]{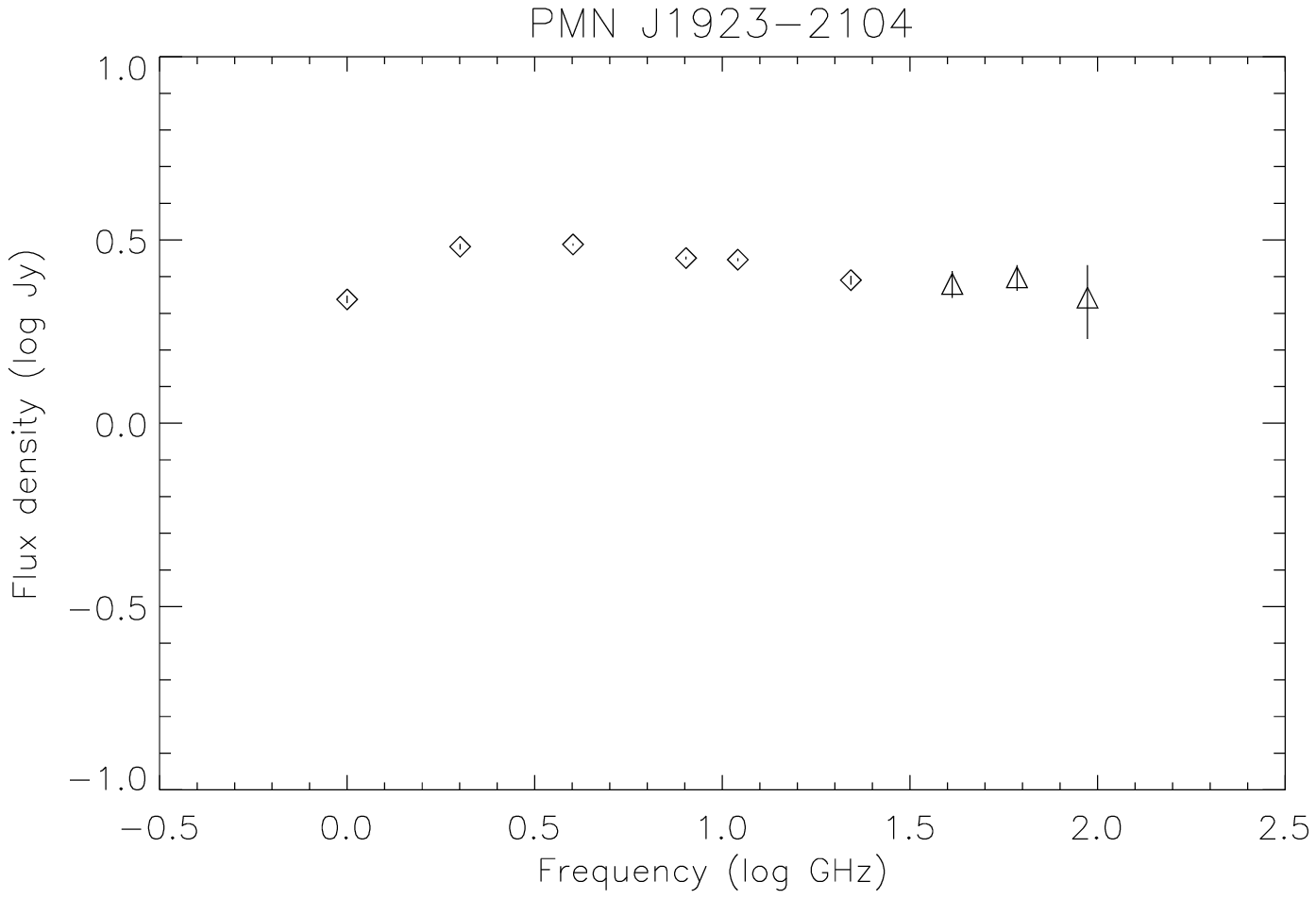}
\includegraphics[width=6cm]{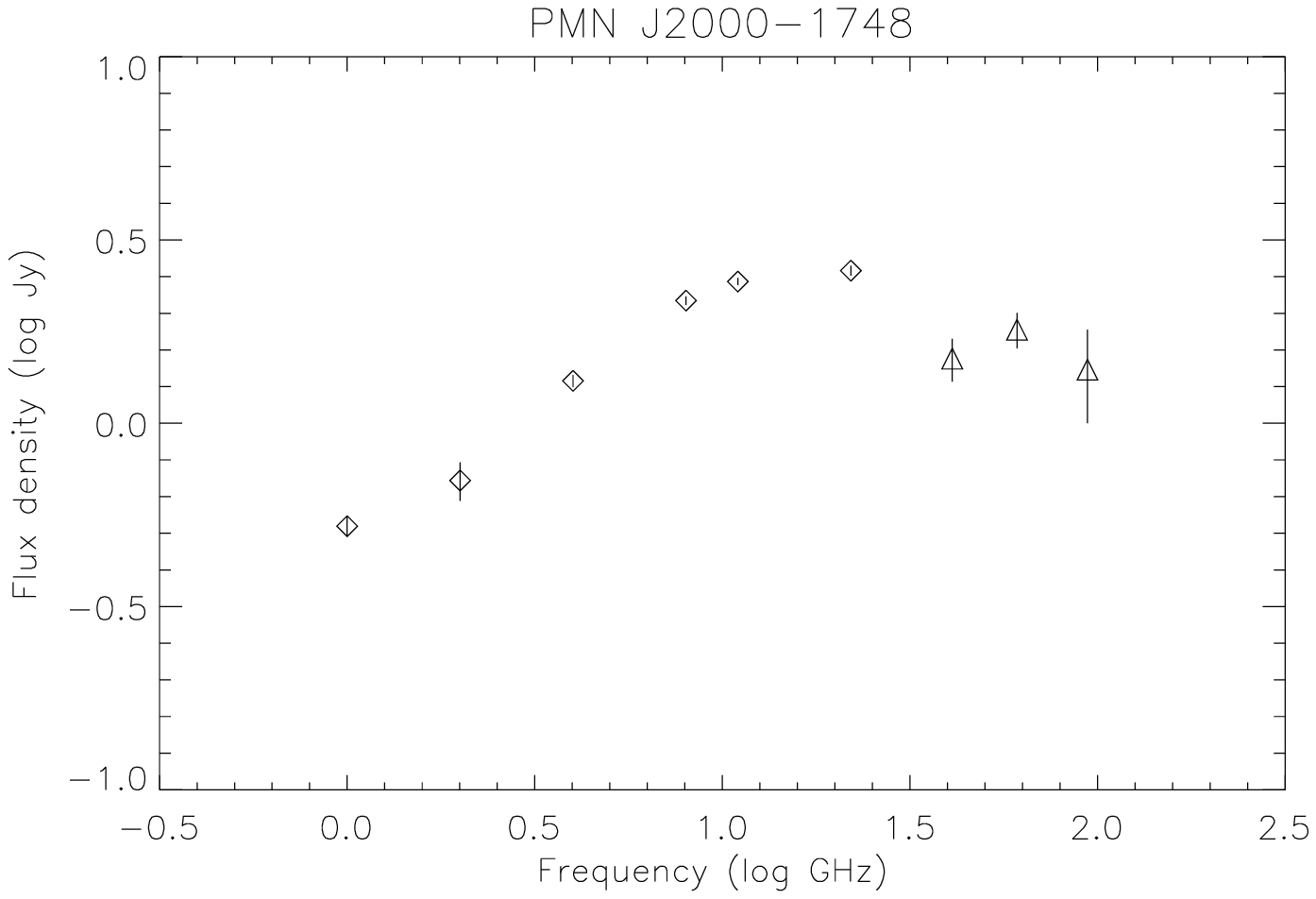}
\end{center}
\caption{ -- Continued}
\end{figure}

 \addtocounter{figure}{-1}
\begin{figure}
\begin{center}
\includegraphics[width=6cm]{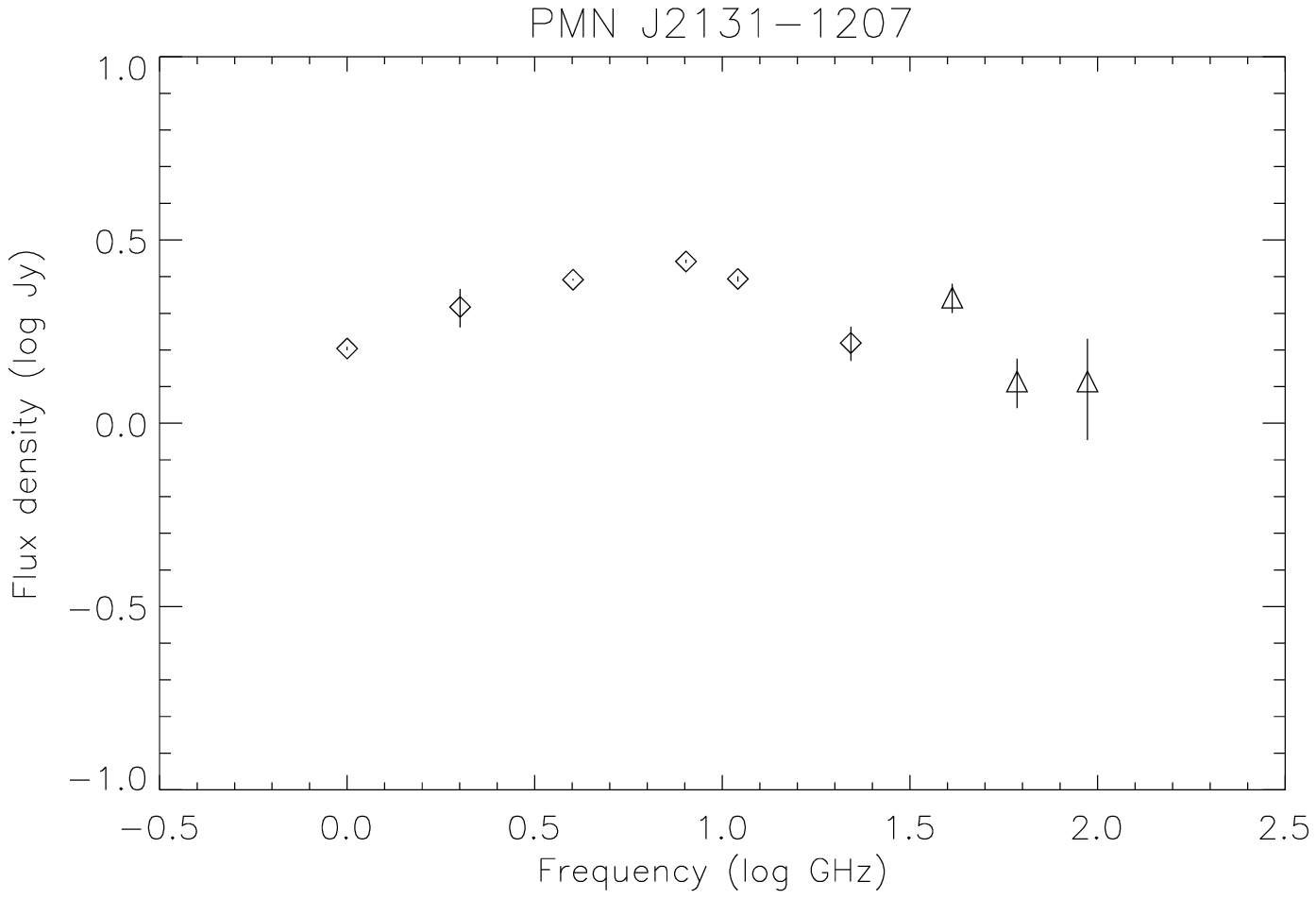}
\includegraphics[width=6cm]{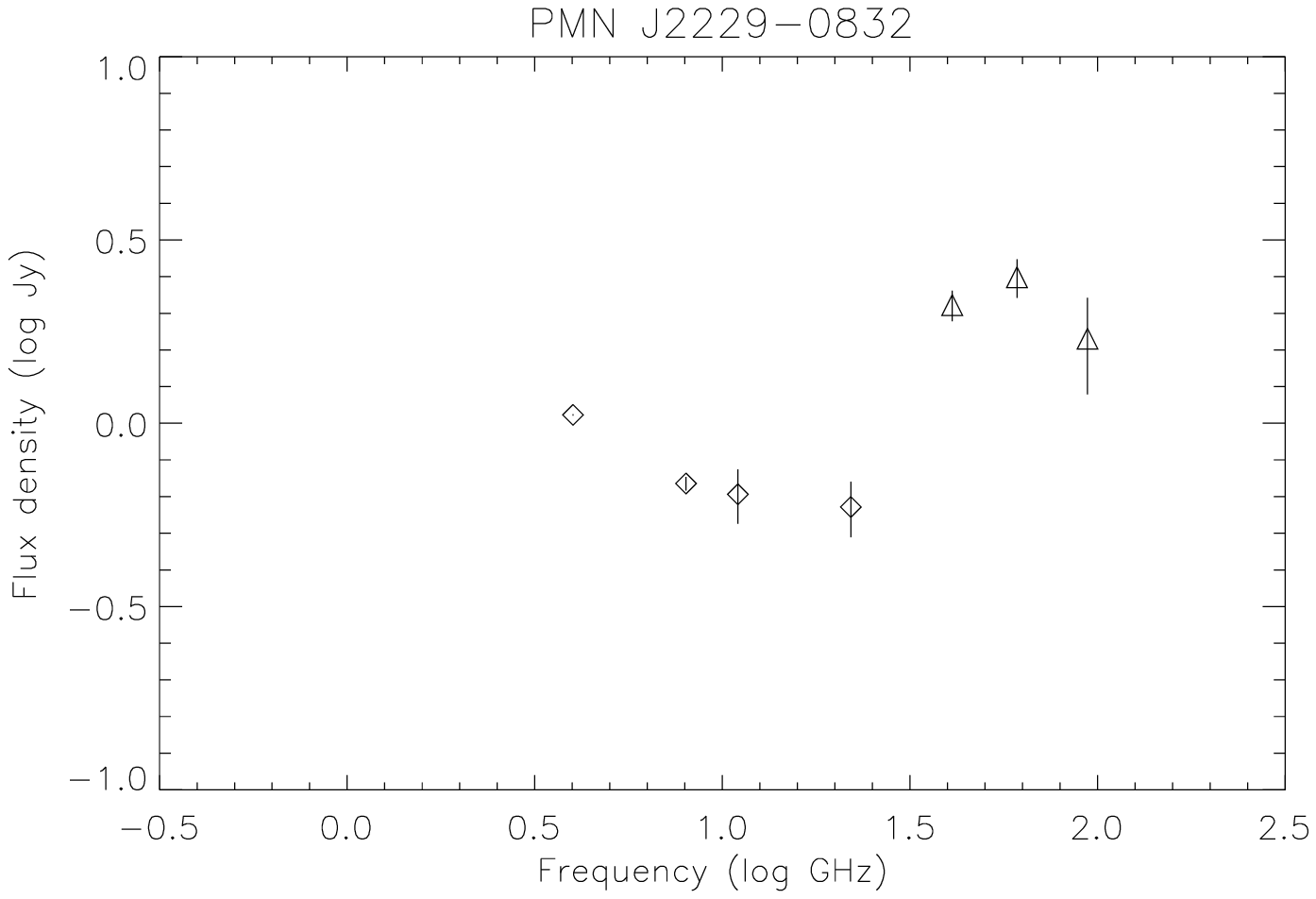}
\end{center}
\caption{ -- Continued}
\end{figure}

\begin{figure}
\begin{center}
\includegraphics[width=6cm]{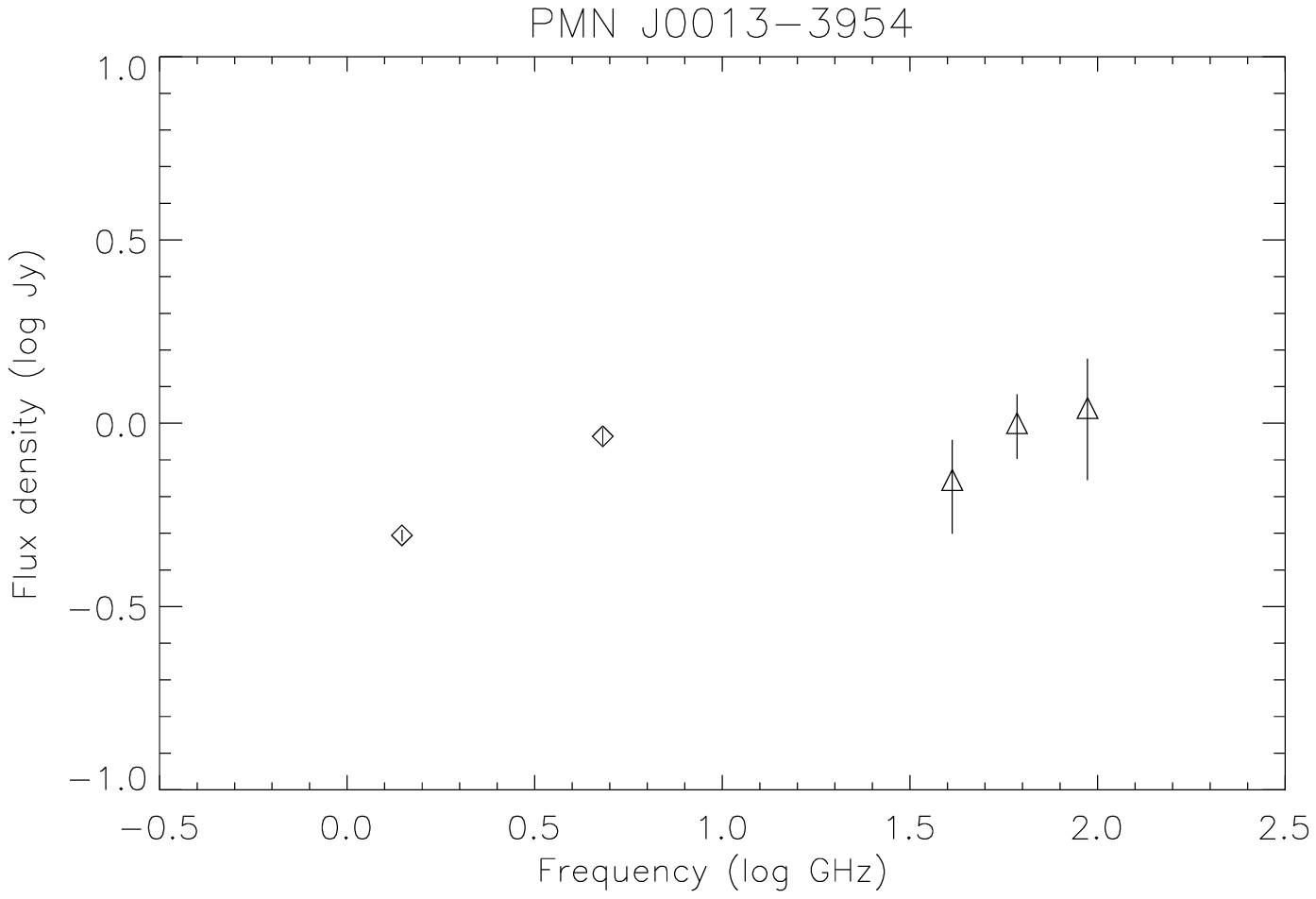}
\includegraphics[width=6cm]{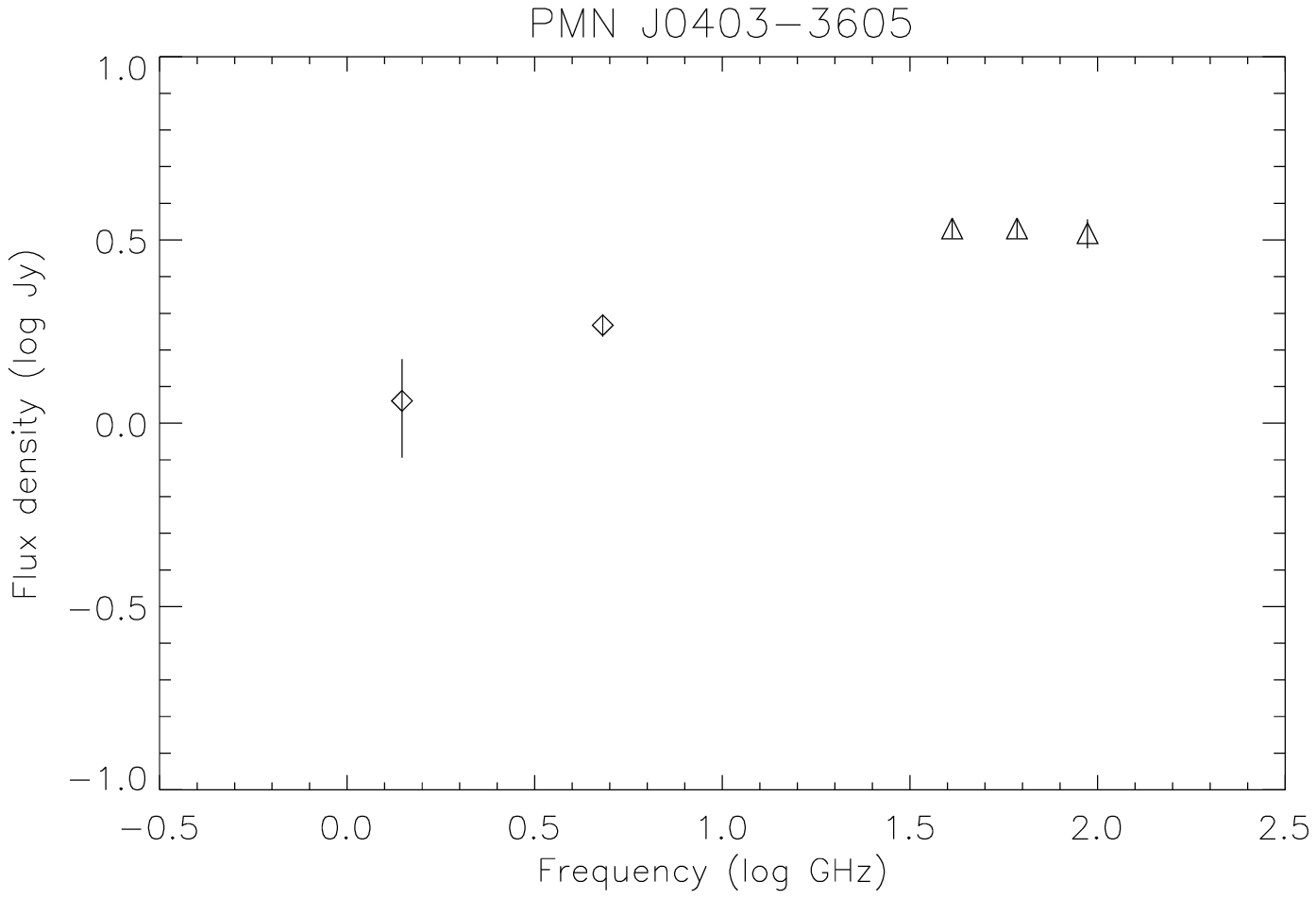}
\includegraphics[width=6cm]{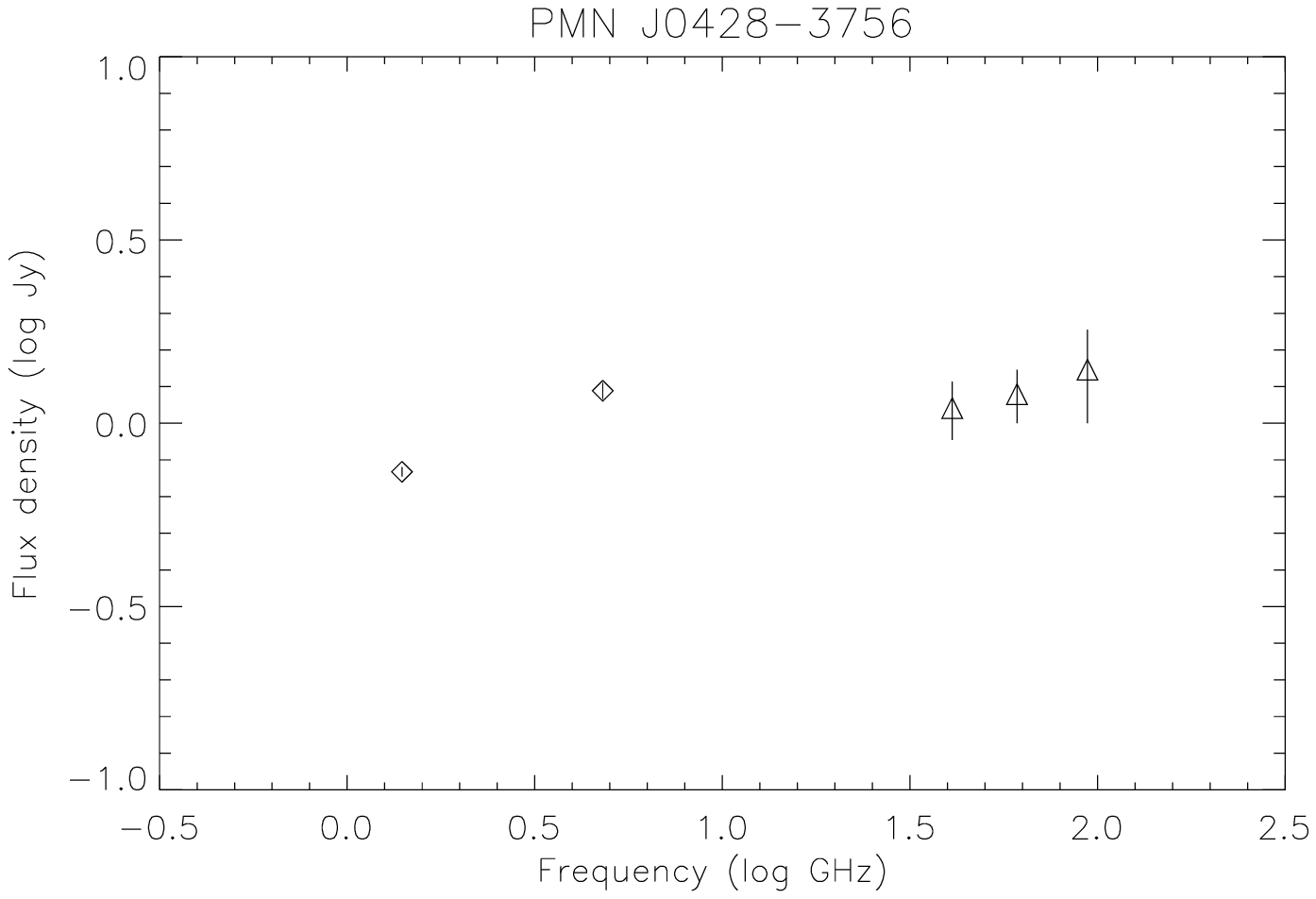}
\includegraphics[width=6cm]{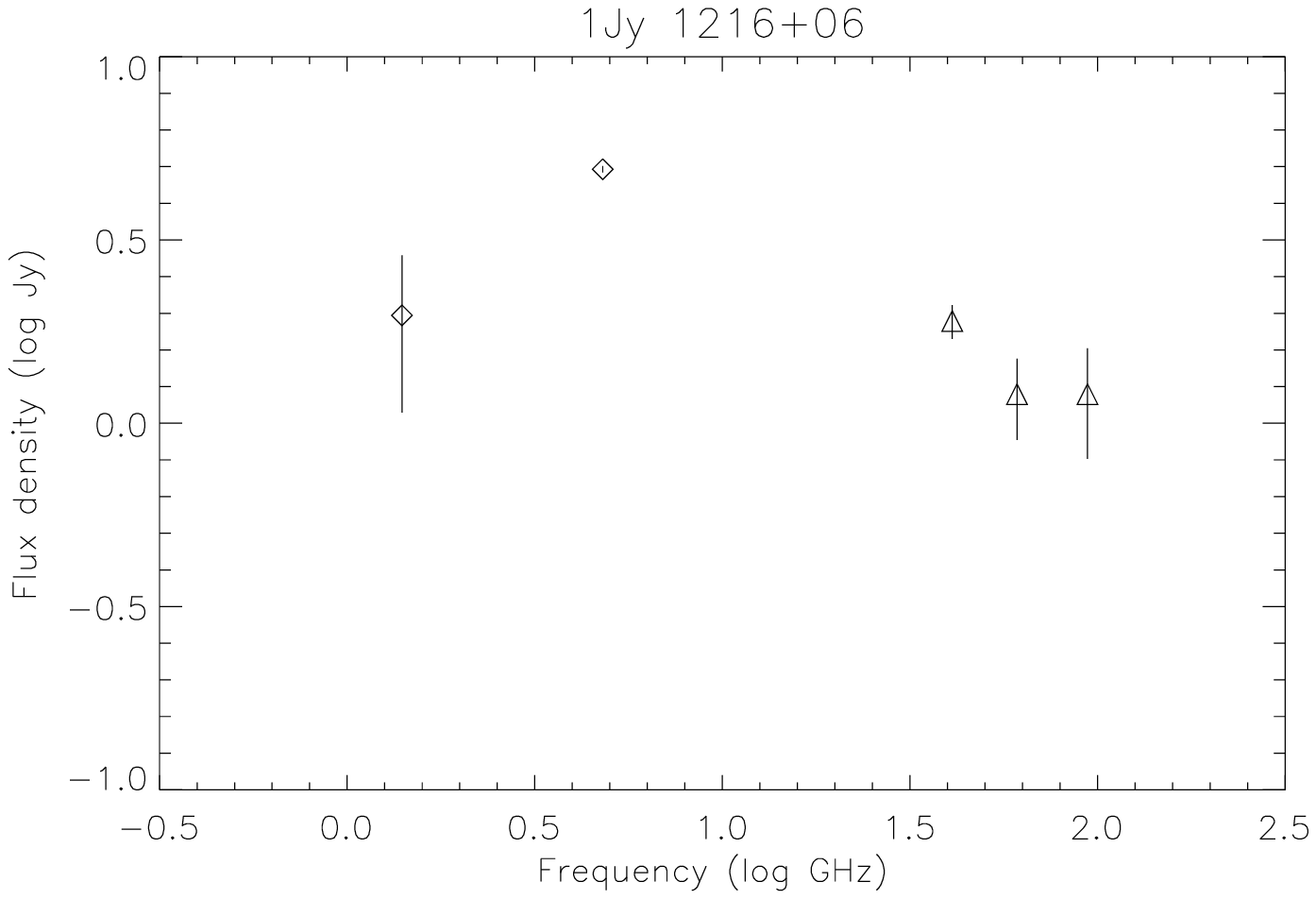}
\includegraphics[width=6cm]{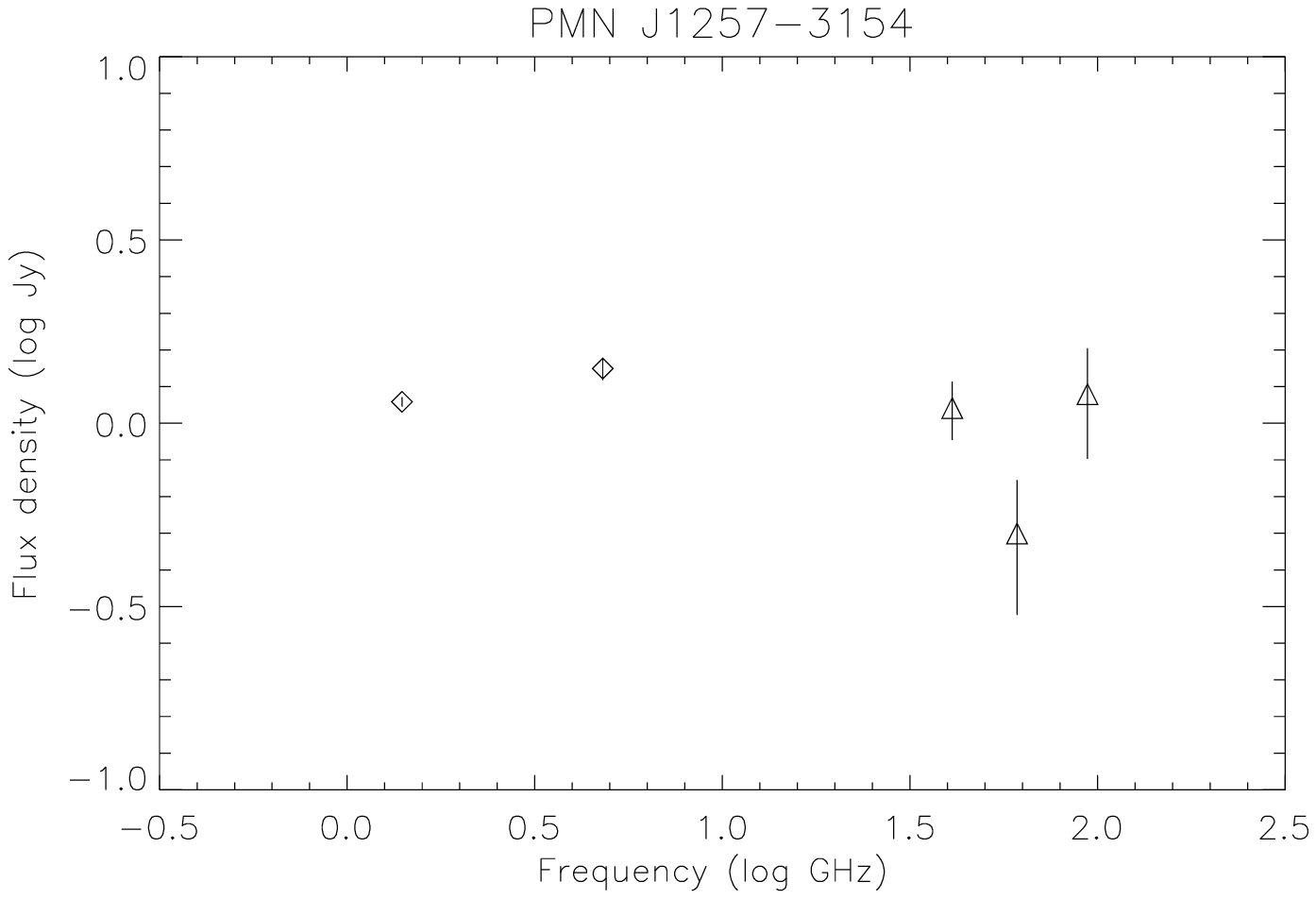}
\includegraphics[width=6cm]{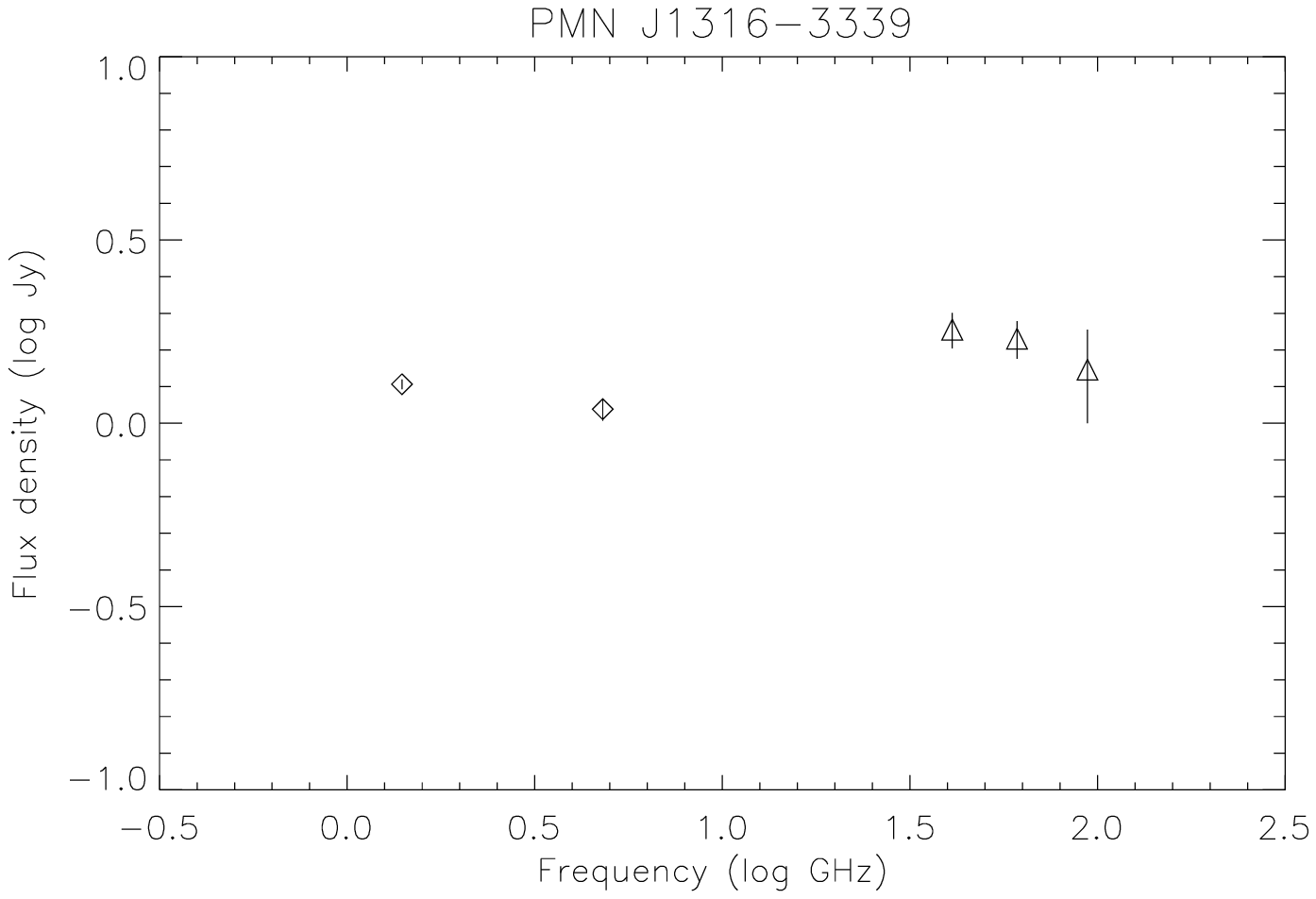}
\includegraphics[width=6cm]{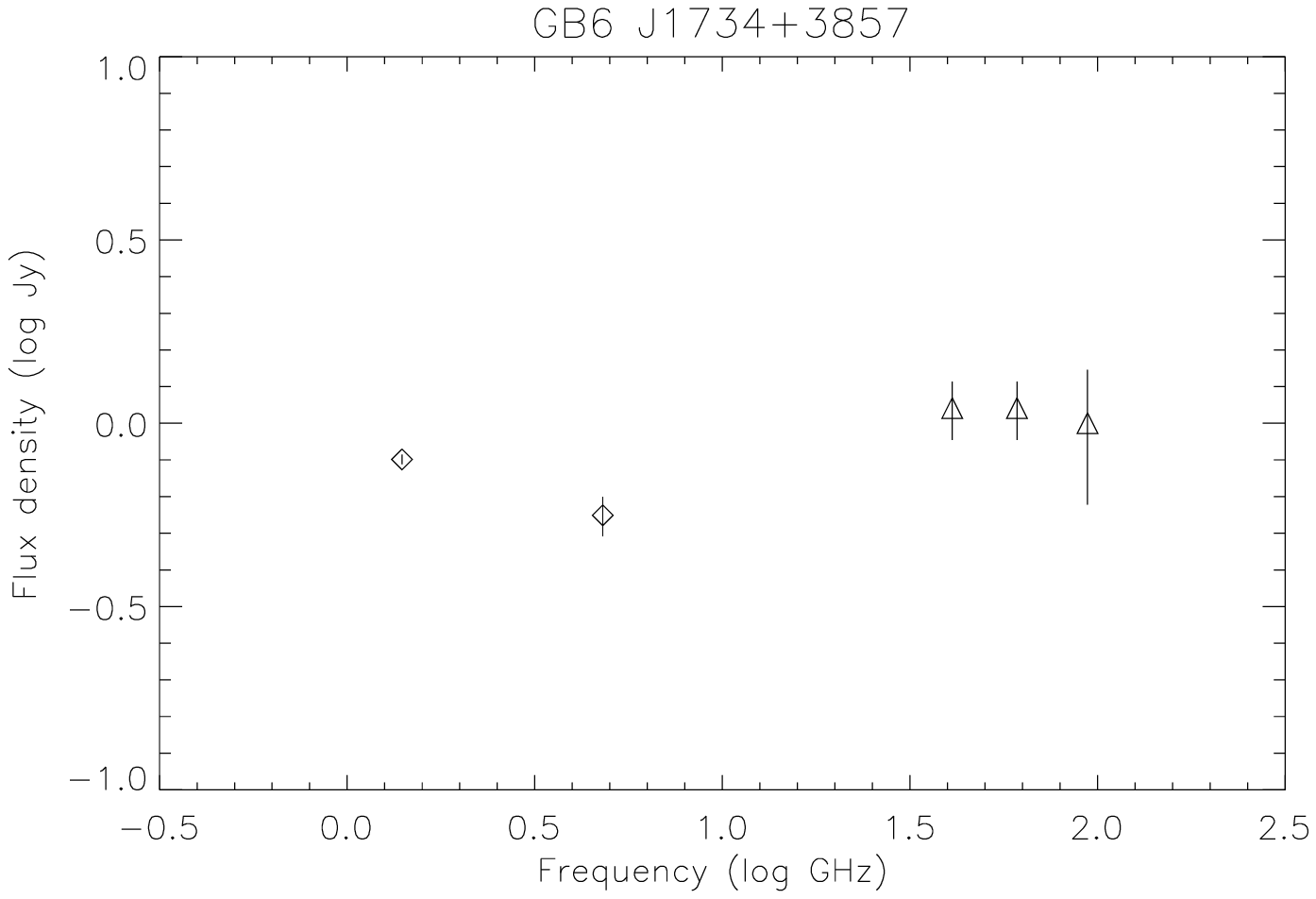}
\includegraphics[width=6cm]{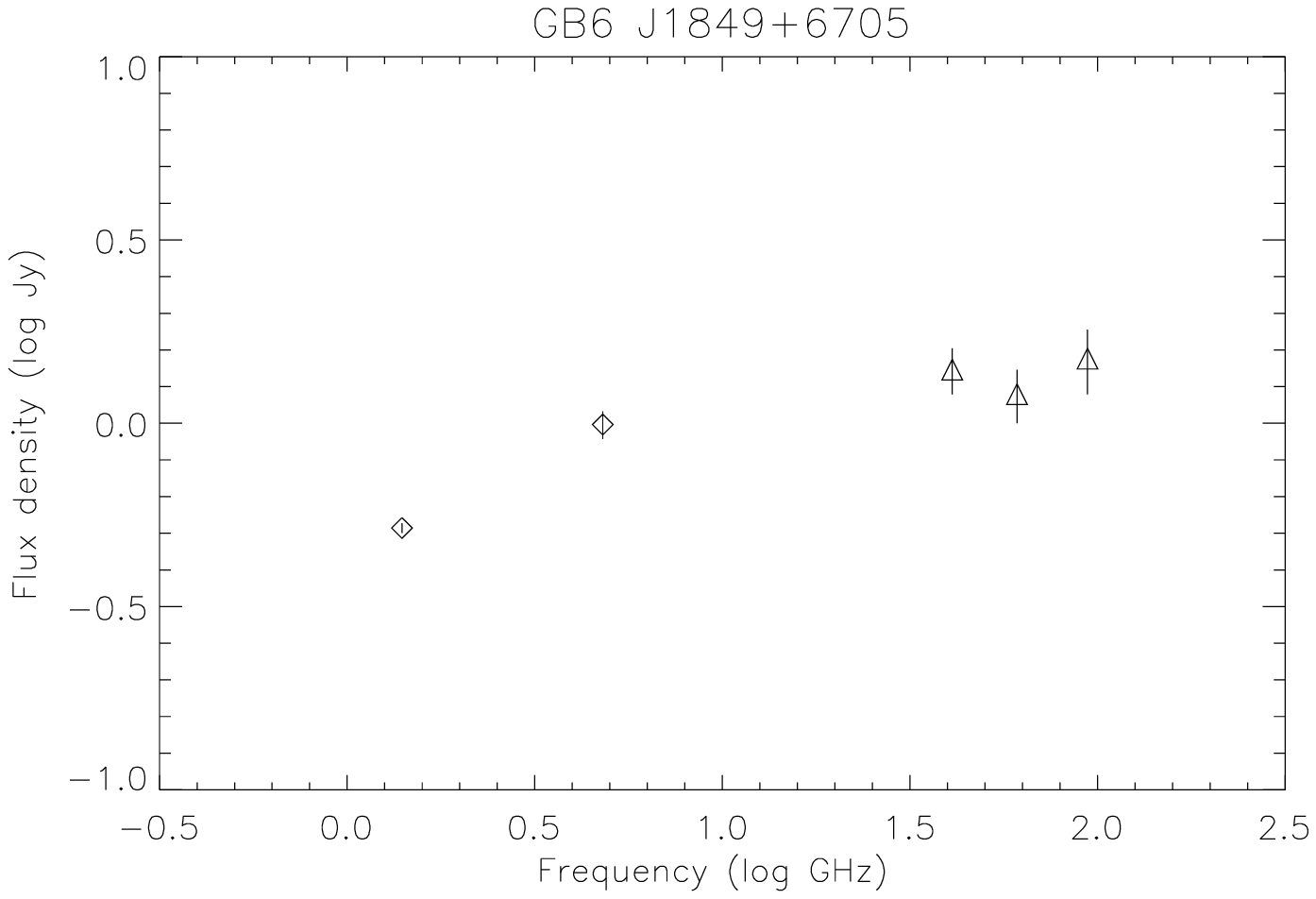}
\end{center}
\caption{Radio spectrum of the sources based on the WMAP catalogue (triangles) and the 1.4 and 4.8 GHz data collected from NED (diamonds)}
\label{spectrumNED} 

\end{figure}

 \addtocounter{figure}{-1}
\begin{figure}
\begin{center}
\includegraphics[width=6cm]{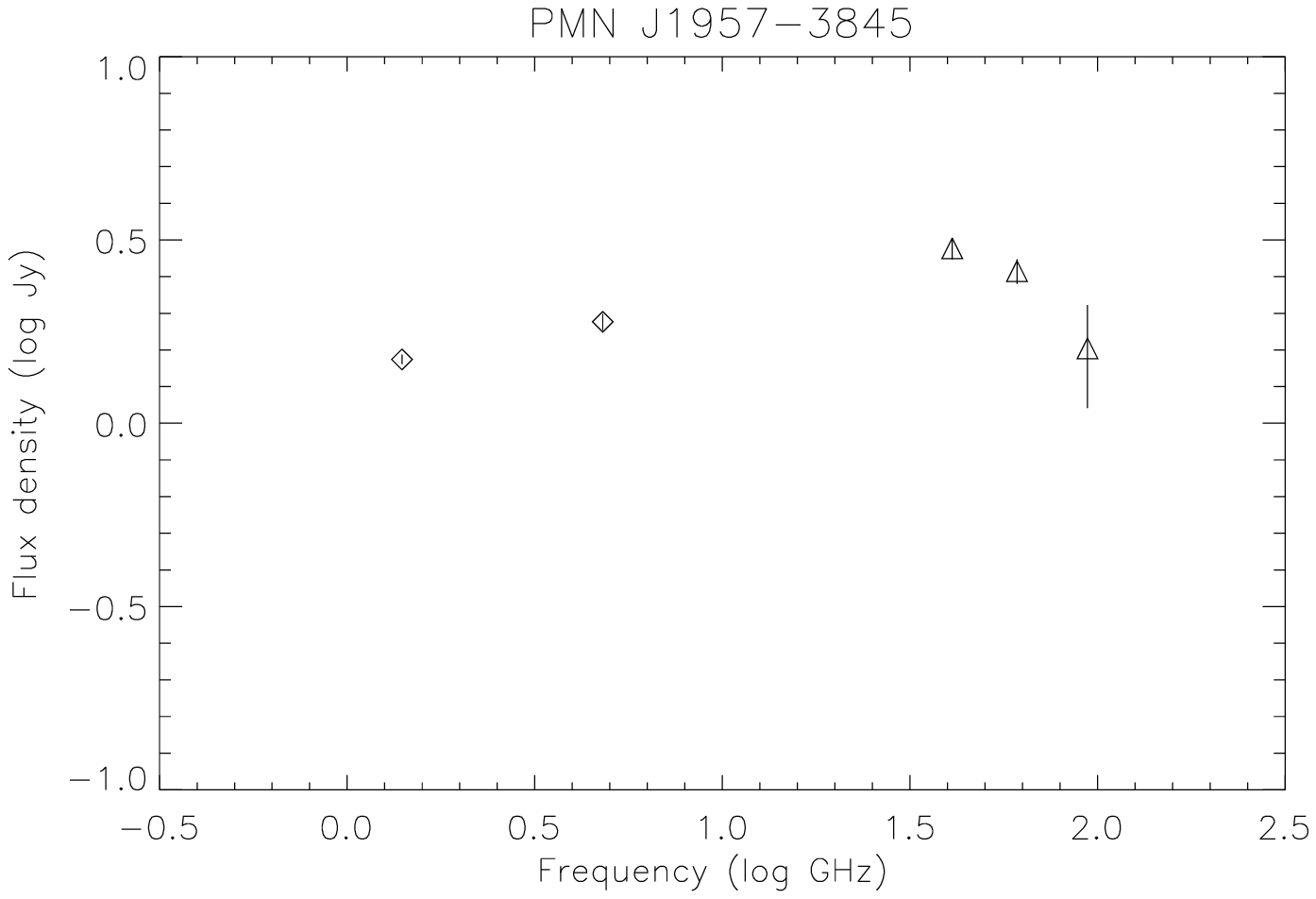}
\includegraphics[width=6cm]{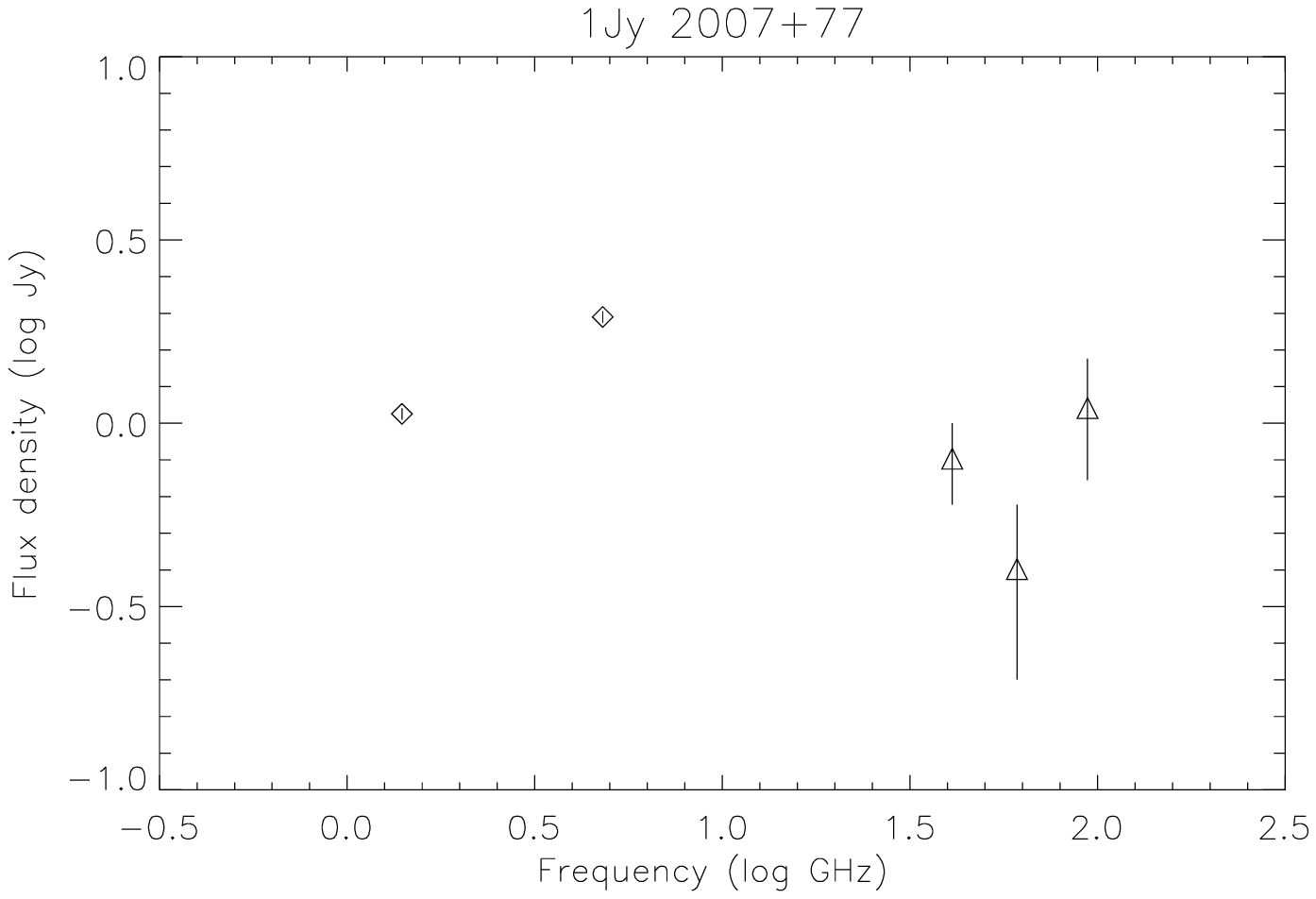}
\includegraphics[width=6cm]{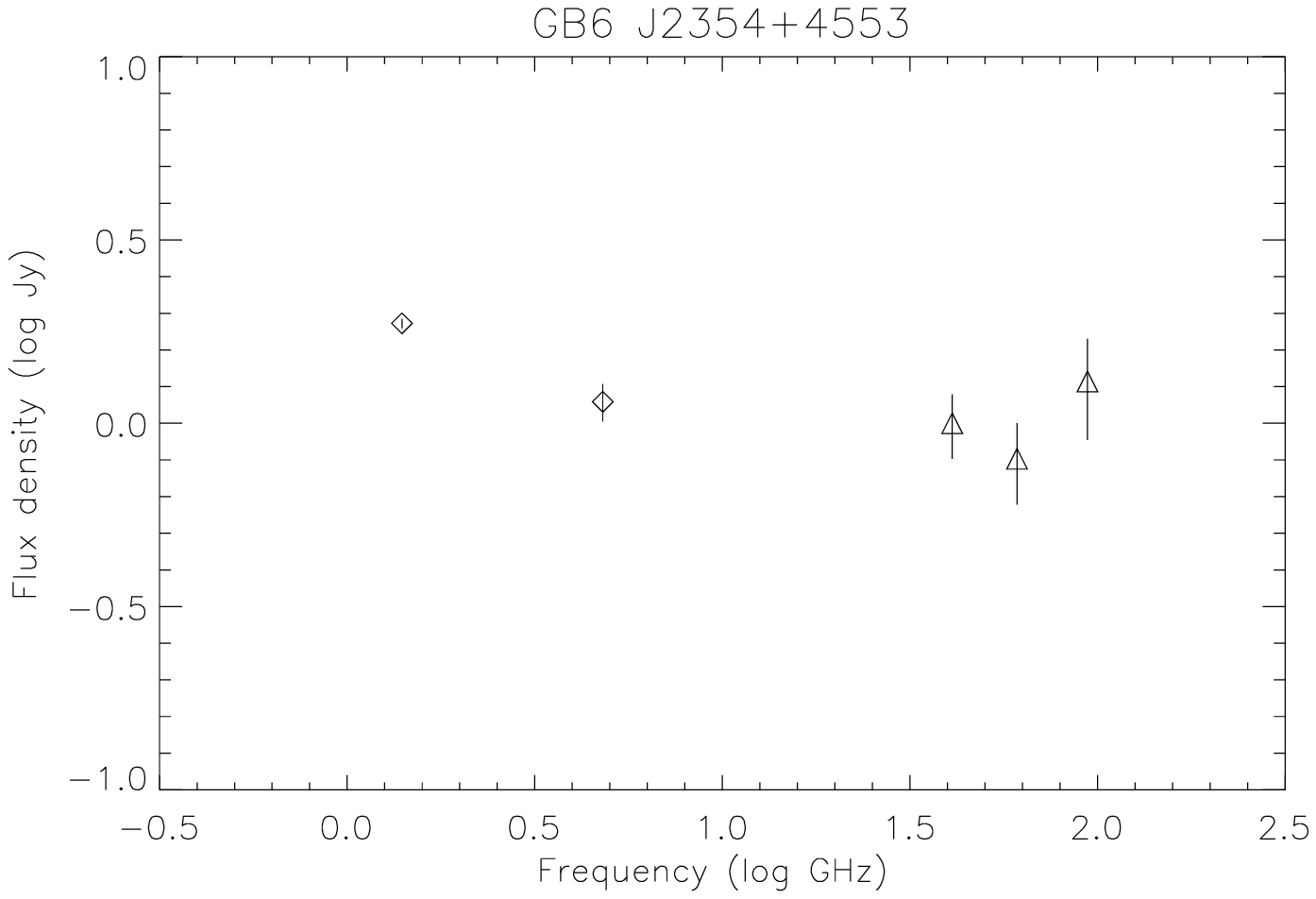}
\end{center}
\caption{ -- Continued}
\label{spectrumNED} 
\end{figure}

\section{Perspectives}

\label{summary}

Using the point source catalogue based on the first 5-year observations
of WMAP \citep{Chen}, we constructed a list of 37 bright extragalactic
sources that are well suited for observations at 86~GHz with ground-based
VLBI, and at 43~GHz with the second-generation ASTRO-G satellite
and its co-observing ground radio telescope network. These sources
have never been imaged with VLBI at 86~GHz, and would extend the
list of known mm-VLBI sources by almost 25\%.

Further improvement is expected from using the point source catalogue
of the Planck spacecraft. The European CMB space mission was launched
in 2009. Throughout its expected lifetime until 2012, it makes sensitive
all-sky measurements at nine different frequencies, from 30 to 857~GHz.
The Planck point source catalogue \citep{Vielva} is estimated to
have a detection limit of $\sim$0.4~Jy at around 100~GHz, and flux
density uncertainties of $\sim15\%$. 
Our method described here could easily be extended to lower flux density levels. 
Thus the Planck list
will be an essential new tool to look for even more potential mm-VLBI
target sources, based on their measured flux densities and spectra.
The most recent Planck Early Release Compact Source Catalog \citep{Ade}
already shows promising perspectives for studying the spectral energy distributions
of extragalactic radio sources. Technical developments and network extensions, including the
addition of the Atacama Large Millimeter and Submillimeter Array (ALMA)
for ground-based mm-VLBI observations, will largely increase the detection
sensitivity in the near future. Regular global VLBI imaging at an
even higher frequency, 230~GHz, will soon be established \citep[e.g.][]{Krichbaum}.

\section*{Acknowledgements}

We are grateful for the constructive suggestions received
from the two anonymous referees. This work was supported by the Hungarian
Space Office of the National Office for Research and Technology (URK09314)
and the Hungarian Scientific Research Fund (OTKA K72515). This research
has made use of the NASA/IPAC Extragalactic Database (NED) which is
operated by the Jet Propulsion Laboratory, California Institute of
Technology, under contract with the National Aeronautics and Space
Administration.

% \appendix

% \bibitem[Names(Year)]{label} or \bibitem[Names(Year)Long names]{label}.
% (\harvarditem{Name}{Year}{label} is also supported.)

\end{document}